\documentclass[12pt]{iopart}
\usepackage{iopams} 
\usepackage{graphics}
\usepackage{pennames}
\usepackage{amssymb}
\usepackage{setstack}
\def\La{\Lambda}
\def\Al{\overline\Lambda}
\def\XI{\Xi^{-}}

\newcommand{\decayarrow}{\makebox[0mm][l]{\rule{0.33em}{0mm}\rule[0.55ex]{0.044em}{1.55ex}}\rightarrow}
\begin{document}
\title[Transverse mass spectra of strange particles in Pb-Pb at 158~$A$~GeV/$c$]
{Study of the transverse mass spectra of strange particles in Pb-Pb 
collisions at 158~$A$~GeV/$c$}
\author{  
The NA57 Collaboration:\\  
F~Antinori$^{l}$, 
P~Bacon$^{e}$, 
A~Badal{\`a}$^{g}$, 
R~Barbera$^{g}$,
A~Belogianni$^{a}$, 
A~Bhasin$^{e}$, 
I~J~Bloodworth$^{e}$, 
M~Bombara$^{i}$, 
G~E~Bruno$^{b}$
\footnote[1]{To
whom correspondence should be addressed (giuseppe.bruno@cern.ch)},
S~A~Bull$^{e}$,
R~Caliandro$^{b}$,
M~Campbell$^{h}$,
W~Carena$^{h}$,
N~Carrer$^{h}$,
R~F~Clarke$^{e}$,
A~Dainese$^{l}$,
A~P~de~Haas$^{s}$,
P~C~de~Rijke$^{s}$,
D~Di~Bari$^{b}$,
S~Di~Liberto$^{o}$,
R~Divi\`a$^{h}$,
D~Elia$^{b}$,
D~Evans$^{e}$,
G~A~Feofilov$^{q}$,
R~A~Fini$^{b}$,
P~Ganoti$^{a}$,
B~Ghidini$^{b}$,
G~Grella$^{p}$,
H~Helstrup$^{d}$,
K~F~Hetland$^{d}$,
A~K~Holme$^{k}$,
A~Jacholkowski$^{g}$,
G~T~Jones$^{e}$,
P~Jovanovic$^{e}$,
A~Jusko$^{e}$,
R~Kamermans$^{s}$,
J~B~Kinson$^{e}$,
K~Knudson$^{h}$,
A~A~Kolozhvari$^{q}$,
V~Kondratiev$^{q}$,
I~Kr\'alik$^{i}$,
A~Krav\v c\'akov\'a$^{j}$,
P~Kuijer$^{s}$,
V~Lenti$^{b}$,
R~Lietava$^{e}$,
G~L\o vh\o iden$^{k}$,
V~Manzari$^{b}$,
G~Martinsk\'a$^{j}$,
M~A~Mazzoni$^{o}$,
F~Meddi$^{o}$,
A~Michalon$^{r}$,
M~Morando$^{l}$,
E~Nappi$^{b}$,
F~Navach$^{b}$,
P~I~Norman$^{e}$,
A~Palmeri$^{g}$,
G~S~Pappalardo$^{g}$,
B~Pastir\v c\'ak$^{i}$,
J~Pi\v s\'ut$^{f}$,
N~Pi\v s\'utov\'a$^{f}$,
R~J~Platt$^{e}$, 
F~Posa$^{b}$,
E~Quercigh$^{l}$,
F~Riggi$^{g}$,
D~R\"ohrich$^{c}$,
G~Romano$^{p}$,
K~\v{S}afa\v{r}\'{\i}k$^{h}$,
L~\v S\'andor$^{i}$,
E~Schillings$^{s}$,
G~Segato$^{l}$,
M~Sen\'e$^{m}$,
R~Sen\'e$^{m}$,
W~Snoeys$^{h}$,
F~Soramel$^{l}$
\footnote[2]{Permanent  
address: University of Udine, Udine, Italy},
M~Spyropoulou-Stassinaki$^{a}$,
P~Staroba$^{n}$,
T~A~Toulina$^{q}$,
R~Turrisi$^{l}$,
T~S~Tveter$^{k}$,
J~Urb\'{a}n$^{j}$,
F~F~Valiev$^{q}$,
A~van~den~Brink$^{s}$,
P~van~de~Ven$^{s}$,
P~Vande~Vyvre$^{h}$,
N~van~Eijndhoven$^{s}$,
J~van~Hunen$^{h}$,
A~Vascotto$^{h}$,
T~Vik$^{k}$,
O~Villalobos~Baillie$^{e}$,
L~Vinogradov$^{q}$,
T~Virgili$^{p}$,
M~F~Votruba$^{e}$,
J~Vrl\'{a}kov\'{a}$^{j}$\ and
P~Z\'{a}vada$^{n}$
}
\address{
$^{a}$ Physics Department, University of Athens, Athens, Greece\\
$^{b}$ Dipartimento IA di Fisica dell'Universit{\`a}
       e del Politecnico di Bari and INFN, Bari, Italy \\
$^{c}$ Fysisk Institutt, Universitetet i Bergen, Bergen, Norway\\
$^{d}$ H{\o}gskolen i Bergen, Bergen, Norway\\
$^{e}$ University of Birmingham, Birmingham, UK\\
$^{f}$ Comenius University, Bratislava, Slovakia\\
$^{g}$ University of Catania and INFN, Catania, Italy\\
$^{h}$ CERN, European Laboratory for Particle Physics, Geneva, Switzerland\\
$^{i}$ Institute of Experimental Physics, Slovak Academy of Science,
       Ko\v{s}ice, Slovakia\\
$^{j}$ P.J. \v{S}af\'{a}rik University, Ko\v{s}ice, Slovakia\\
$^{k}$ Fysisk Institutt, Universitetet i Oslo, Oslo, Norway\\
$^{l}$ University of Padua and INFN, Padua, Italy\\
$^{m}$ Coll\`ege de France, Paris, France\\
$^{n}$ Institute of Physics, Prague, Czech Republic\\
$^{o}$ University ``La Sapienza'' and INFN, Rome, Italy\\
$^{p}$ Dipartimento di Scienze Fisiche ``E.R. Caianiello''
       dell'Universit{\`a} and INFN, Salerno, Italy\\
$^{q}$ State University of St. Petersburg, St. Petersburg, Russia\\
$^{r}$ IReS/ULP, Strasbourg, France\\
$^{s}$ Utrecht University and NIKHEF, Utrecht, The Netherlands
}
\begin{abstract}
The NA57 experiment has collected high statistics, high purity samples 
of \PKzS\ and \PgL, $\Xi$\ and  $\Omega$\ hyperons produced in Pb-Pb 
collisions at 158 $A$\ GeV/$c$. 
In this paper we present a study of the transverse mass spectra of 
these particles for a  sample of events corresponding to the 
most central 53\% of the inelastic Pb-Pb cross-section. 
We analyse the transverse mass distributions in the framework 
of the blast-wave model for the full sample  
and, for the first  time at the SPS, as a function of the event 
centrality.
\end{abstract}


\submitto{\JPG}

\maketitle

\section{Introduction} 
The study of ultrarelativistic heavy-ion collisions is motivated mainly by 
the quantum chromodynamics (QCD) prediction that at sufficiently high energy   
density the excited nuclear matter undergoes a phase transition into a 
system of deconfined quarks and gluons (quark-gluon plasma, QGP)~\cite{Cab}.  

Strange particles have proved  
over the past years to be a powerful tool for the study of reaction dynamics 
in high-energy heavy-ion collisions.  
\newline
The WA97 experiment has measured~\cite{WA97Enh}  
an enhanced production of particles carrying one, two or three units  
of strangeness in central Pb-Pb collisions at 158 $A$\ GeV/$c$\  
with respect to proton  
induced reactions (strangeness enhancement). The observed pattern of the  
enhancements as a function of strangeness 
was predicted more than 20 years ago  
to be a consequence of a QGP formation~\cite{Rafelski}. 

NA57 is a dedicated 
experiment at the CERN SPS 
for the study of the production of 
strange and multi-strange particles  
in Pb-Pb  collisions~\cite{NA57proposal}.   
It continues and extends the study initiated by its predecessor WA97, 
by {\em (i)} enlarging the triggered fraction of the inelastic  
cross-section thus extending the centrality range towards less central  
collisions and {\em (ii)} collecting data also at lower (40 $A$\ GeV/$c$)  
beam momentum in order to study the energy dependence of the 
strangeness enhancements.

In this paper we concentrate on the analysis of the transverse mass 
($m_{\tt T}=\sqrt{p_{\tt T}^2+m^2}$)    
spectra for 
\PgL, \PgXm, \PgOm\ hyperons, their antiparticles and \PKzS\   
measured in Pb-Pb collisions at 158 $A$\ GeV/$c$.  

The $m_{\tt T}$\  
spectra are expected to be sensitive to the details of the production  
dynamics~\cite{BlastRef,BlastRef2}.  
In particular, for a fireball in local thermal equilibrium, 
the shapes of the $m_{\tt T}$\ spectra depend both on the thermal motion of the 
particles and on the collective flow driven by the pressure. 
To disentangle the two contributions, namely 
thermal motion and transverse flow, one has to rely on models. 
Several freeze-out scenarios 
have been proposed (for an 
overview refer to~\cite{Torrieri}). 
They differ in the freeze-out geometry and in 
the flow velocity profile. In the present analysis we consider the 
{\it blast-wave} model~\cite{BlastRef,BlastRef2},  
which assumes cylindrical symmetry for an expanding fireball in local 
thermal equilibrium, with different hypotheses on the transverse flow profile. 

The paper is organized as follows. The NA57 apparatus is briefly 
described in section 2. Section 3 deals with the measurements  
of the transverse mass spectra.  
Inverse slope parameters are presented and discussed in section 4.   
The blast-wave model is introduced in section 5 
and compared to data under different conditions. 
Finally, conclusions are drawn in section 6.  
\section{The NA57 experiment}
\begin{figure}[hbt]
\centering
\resizebox{0.78\textwidth}{!}{%
\includegraphics{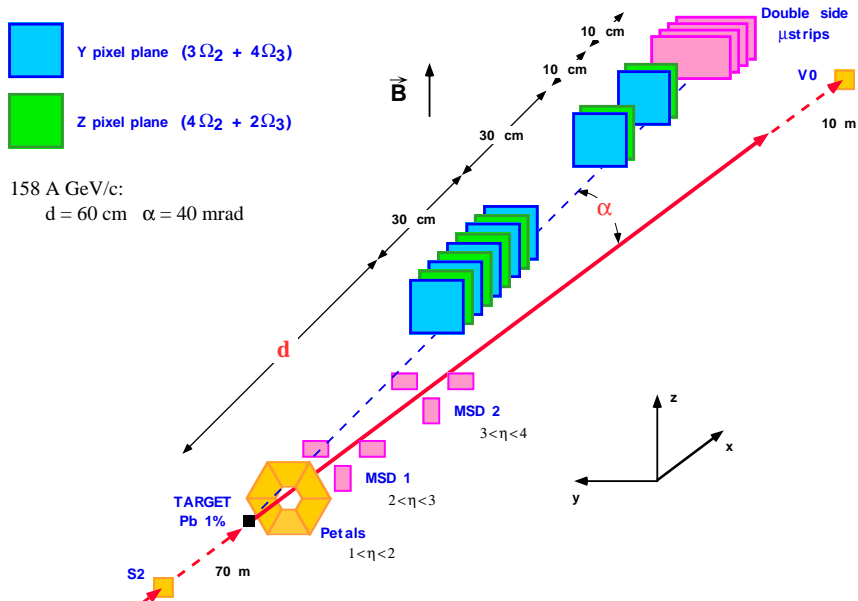}}
\caption{\rm The NA57 apparatus placed inside the 1.4 T field of the GOLIATH 
 magnet.} 
\label{fig:setup}
\end{figure}

The NA57 apparatus, schematically shown in figure~\ref{fig:setup},  
has been described in detail elsewhere~\cite{MANZ}.  
The \PgL, \PgXm, \PgOm\ hyperons, their antiparticles and the \PKzS\ mesons  
are identified by reconstructing their weak  
decays into final states containing only charged particles: 
\begin{equation}
\label{eq:decay}
\begin{array}{lllllll}
    \PKzS &\rightarrow & \pi^+ + \pi^-  & \hspace{10mm}&
    \La  &\rightarrow & p + \pi^-                             \\ \\
    \XI &\rightarrow & \La + \pi^- &  \hspace{10mm}&  \Omega^- &\rightarrow &
   \La + K^-  \\
            &            &
   \decayarrow  p + \pi^-  &  &  &  & \decayarrow  p + \pi^-   \\
\end{array}
\end{equation}
and the corresponding charge conjugates for antihyperons.   
\newline  
Tracks are measured in the silicon tracking 
telescope, an array of pixel detector planes with  
$5 \times 5$\ ${\rm cm}^2$\ cross-section, with a total length 
of about 30 cm. To improve the momentum resolution of high momentum tracks a     
lever arm detector --- an array of additional pixel planes and 
double-sided silicon microstrip detectors ---  
is placed downstream of the tracking telescope.
The whole silicon telescope is placed in 
a 1.4 Tesla magnetic field,
above the beam line, inclined and aligned
with the lower edge of the detectors laying on a line pointing back to the target.
The inclination angle $\alpha$\ with respect to the beam line
and the distance $d$\ of the first pixel plane from the target are 
set  
to accept particles produced in about half a unit of rapidity
around central rapidity and medium transverse momentum: 
$\alpha=40$\ mrad and $d=60$\ cm.  

\noindent
An array of 
scintillation counters (Petals), placed  10 cm downstream of the target,
provides a fast signal to trigger on the centrality of the collisions.  
The triggered fraction of the Pb-Pb inelastic cross-section is about 60\%.  
The centrality of the Pb-Pb collisions is determined (off-line) by analyzing the 
charged particle multiplicity measured by two stations of silicon strip 
detectors (MSD)  
which sample the pseudorapidity intervals $2<\eta<3$\ and $3<\eta<4$.  
\section{Data analysis}
\begin{figure}[hbt]
\centering
\resizebox{0.65\textwidth}{!}{%
\includegraphics{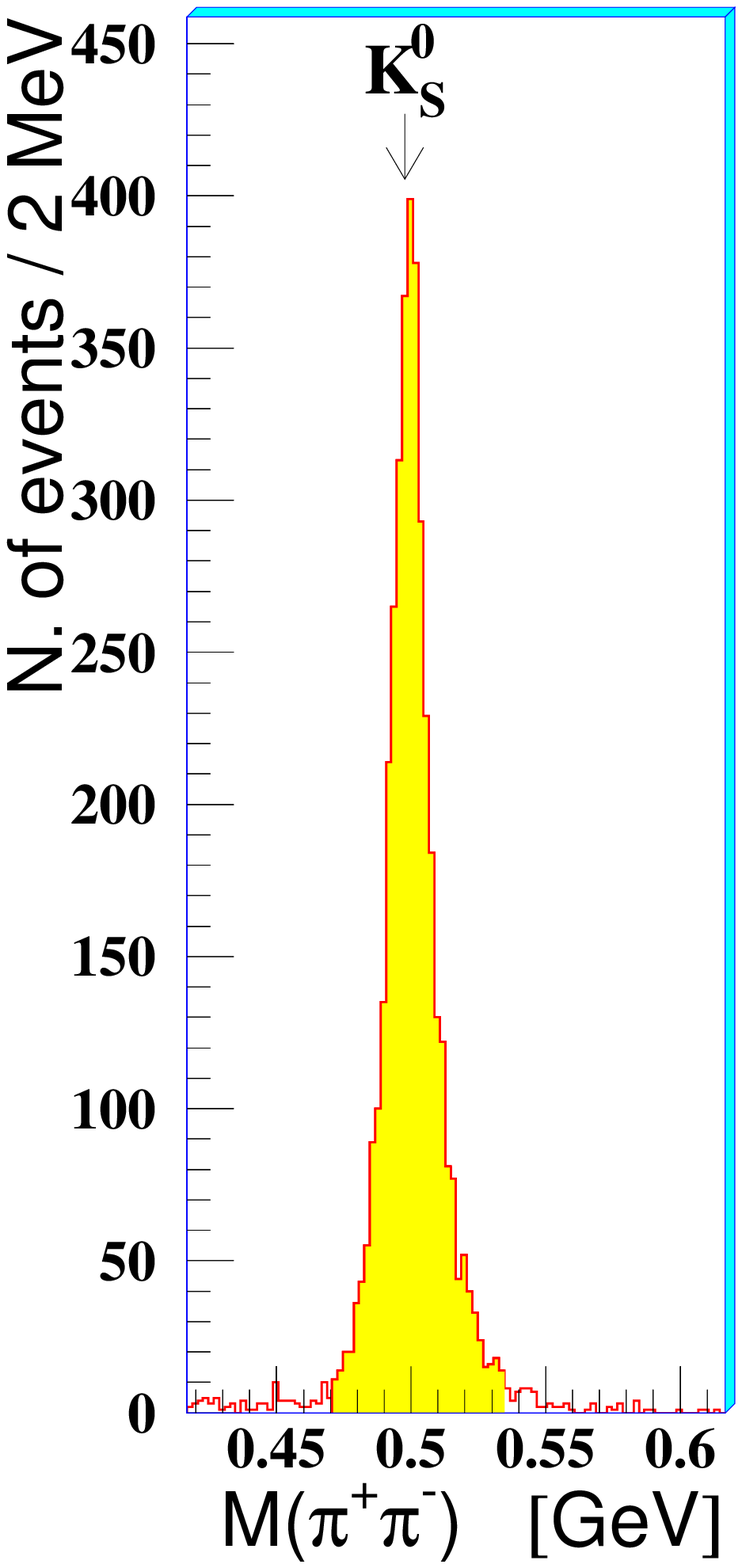}
\includegraphics{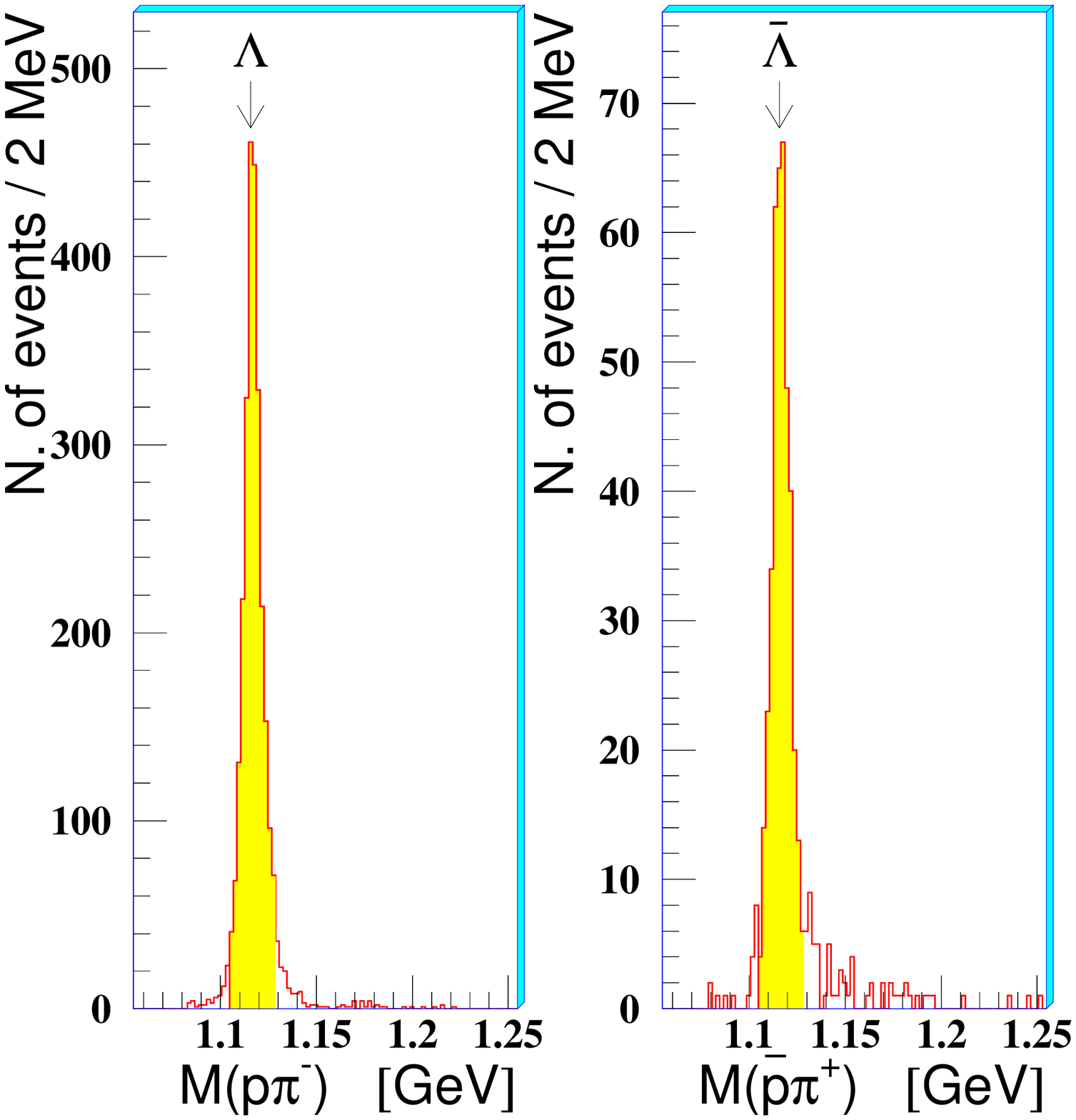}}\\
\resizebox{0.84\textwidth}{!}{%
\includegraphics{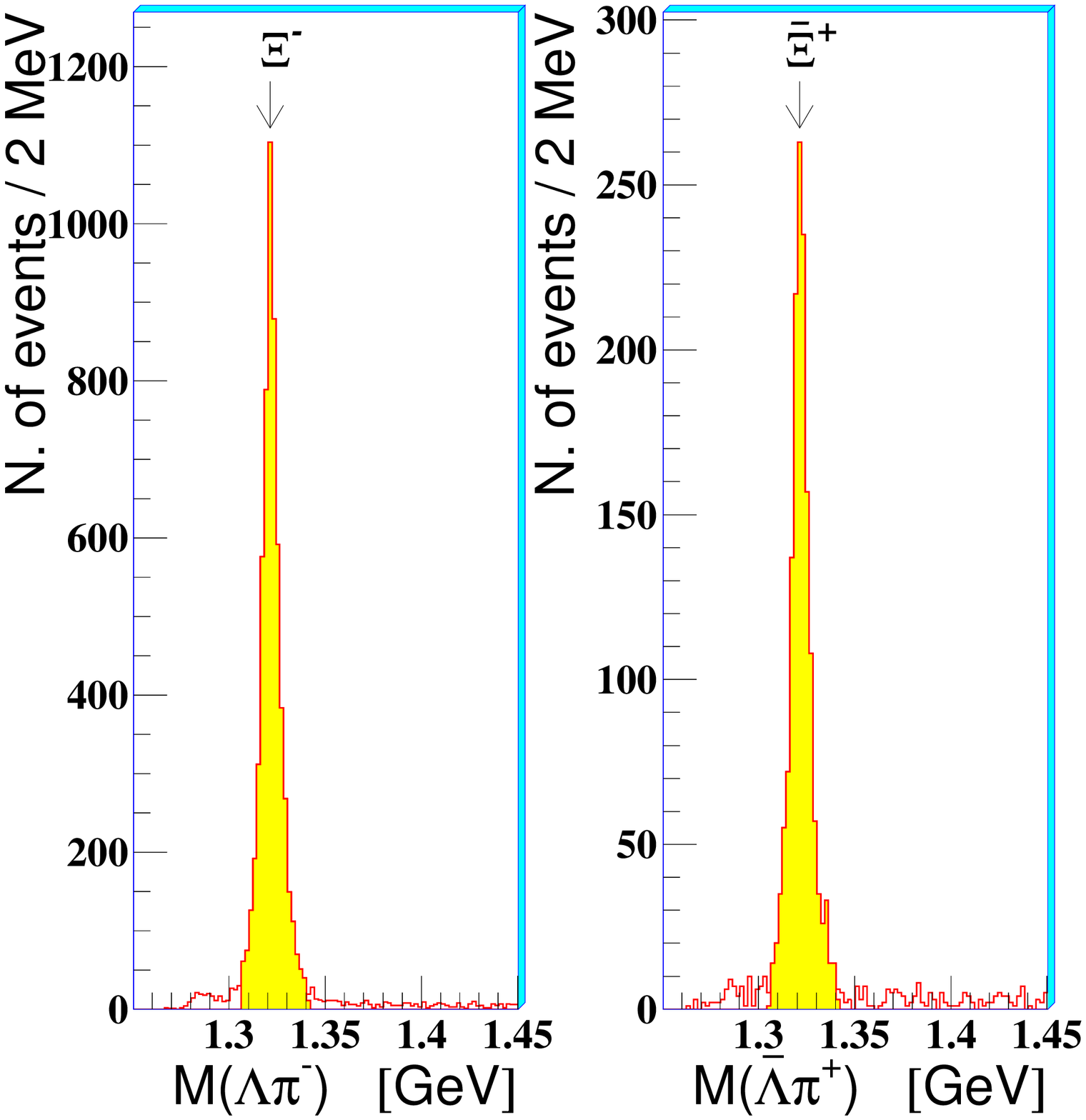}
\includegraphics{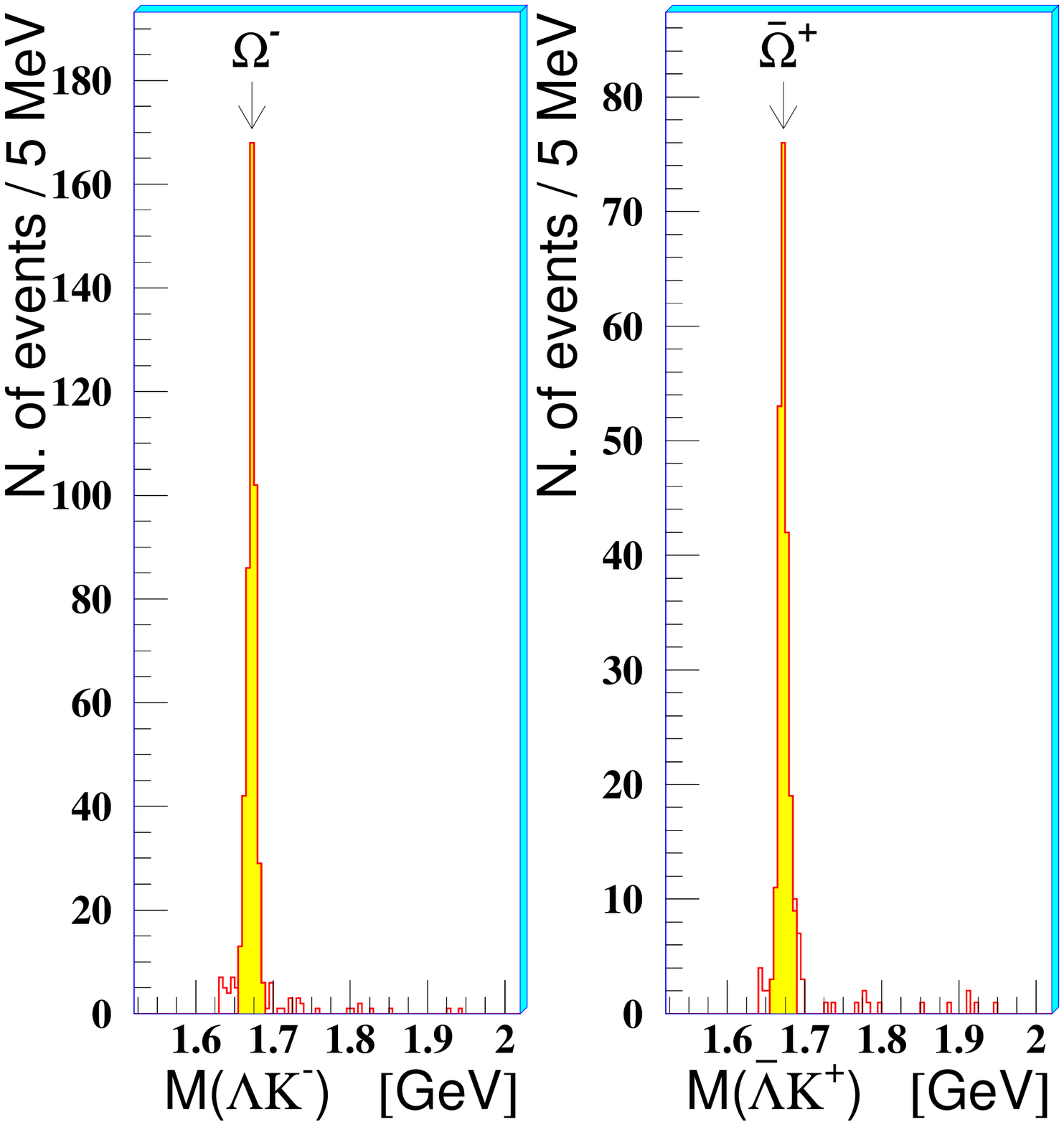}}
\caption{\rm Sample invariant mass spectra for \Pgpp\Pgpm, \Pp\Pgp,  
              \PgL\Pgp\ and \PgL\PK.}
\label{fig:signals}
\end{figure}

The results presented in this paper are based on the analysis of the full data  
sample consisting of 460 M events of Pb-Pb collisions. 
\newline
The strange particle signals are extracted using geometric and kinematic 
constraints, with a method similar to that used in the WA97 
experiment~\cite{WA97PhysLettB433}. As an example of the quality of the 
NA57 data, the invariant mass spectra for \Pgpp\Pgpm, \Pp\Pgp,  \PgL\Pgp\ and 
\PgL\PK\ after all analysis cuts are shown in figure~\ref{fig:signals}.  
Hyperon and \PKzS\ 
peaks are well centered at the PDG values~\cite{PDG} 
with FWHM of 
about 10 MeV (15 MeV for the \PKzS).  

\begin{figure}[hbt]
\centering
\resizebox{0.77\textwidth}{!}{%
\includegraphics{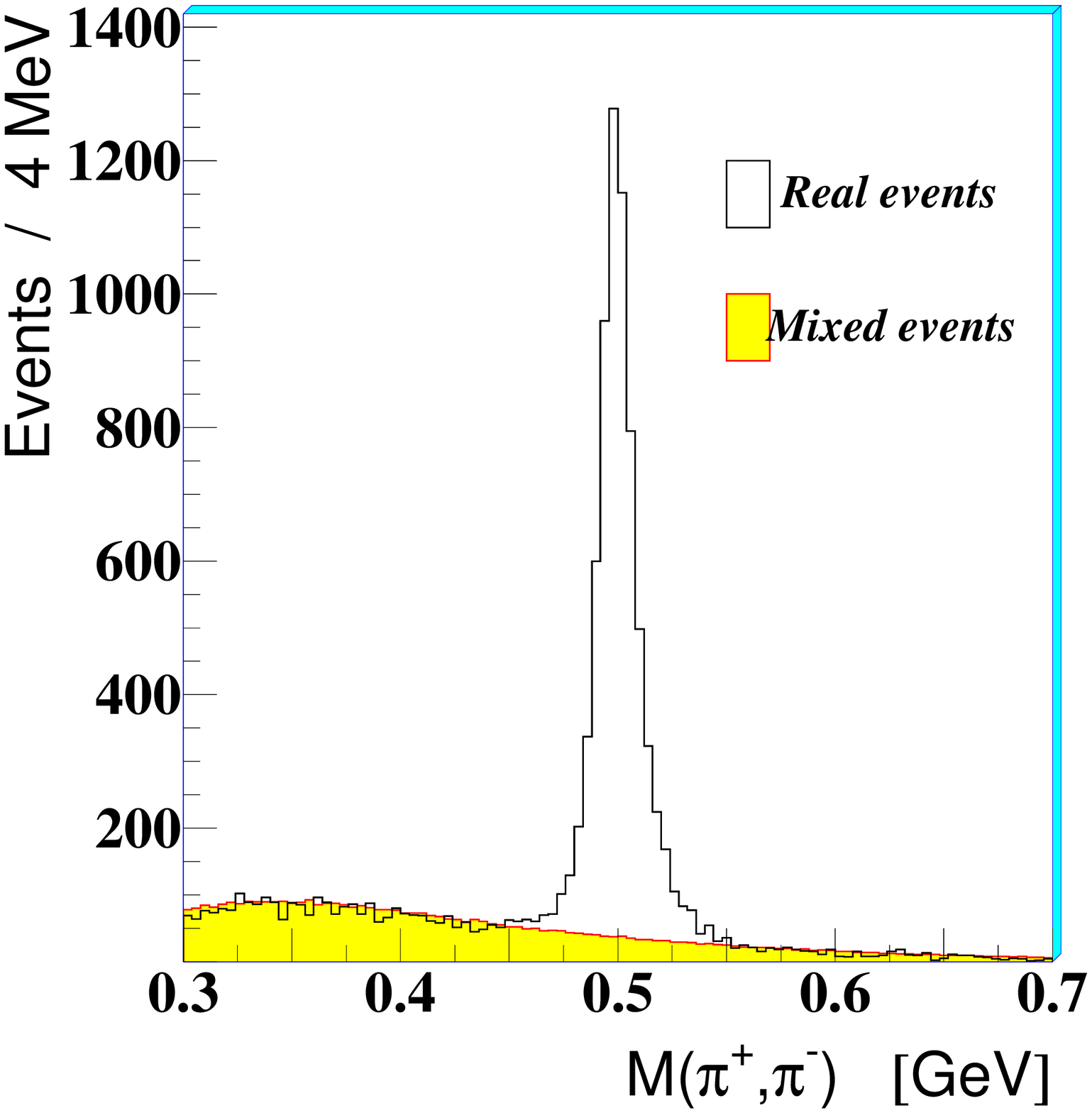}
\includegraphics{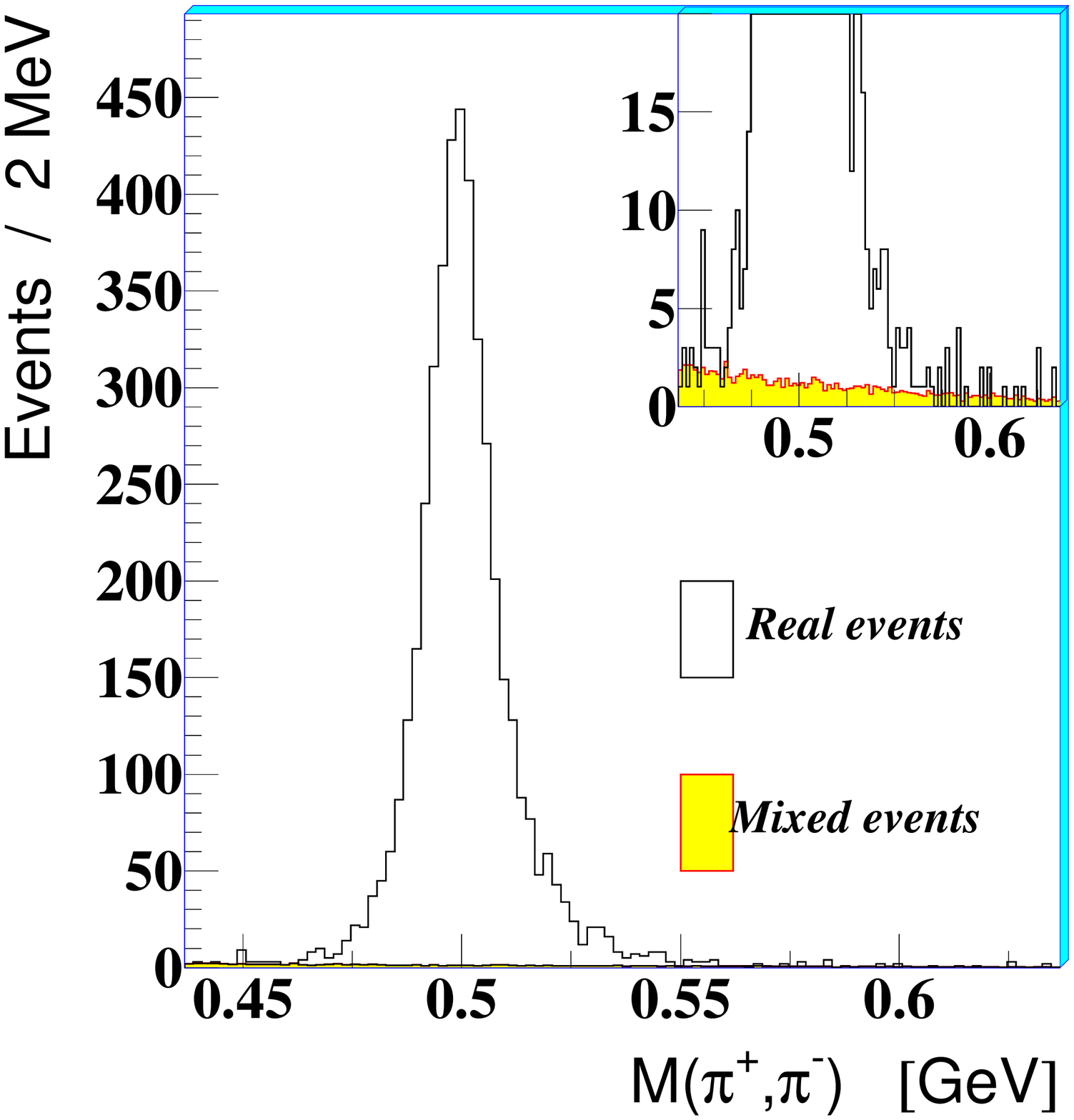}}
\caption{\rm 
The \Pgpp\Pgpm\ invariant mass distribution  
for real and mixed events 
before analysis cuts (left) and after analysis cuts (right).  
The inset in the right plot is a zoom on the vertical axis.}
\label{fig:Mixing}
\end{figure}

Although the mass spectra show very low background, a detailed study of the 
residual combinatorial background has been performed for the high statistics 
samples (\PKzS, \PgL\ and \PagL) 
using 
the method of {\em event mixing}. Fake \PgL, \PagL\ or \PKzS\ ($V^0$) candidates   
are built by pairing all the negative particles from one event  
with all the positive ones from a different event, selecting events 
which are close in multiplicity. 
Then the fake $V^0$s from mixed events are processed with 
reconstruction and analysis programs, like real events.  
In this way we obtain a good description of the combinatorial background.  
The absolute normalization is fixed by the number  
of pairs of oppositely charged particles in real and mixed events.  
Figure~\ref{fig:Mixing} shows the \Pgpp\Pgpm\ invariant mass distribution for
real and mixed events before (left) and  after (right)  
the application of the analysis cuts.  
The total amount of combinatorial background is estimated to be	 
 0.7\%, 0.3\% and 1.2\% for \PKzS, \PgL\ and \PagL, respectively.  
We therefore neglect it.  
Additional details of this method can be found in reference~\cite{BrunoMoriond02}. 
\newline
The estimates of the $\Xi$\ and $\Omega$\ background,  
evaluated with a similar technique, are less than 4\% and 6\%,   
respectively. They have also been neglected for this analysis.   

\begin{figure}[hbt]
\centering
\resizebox{0.70\textwidth}{!}{%
\includegraphics{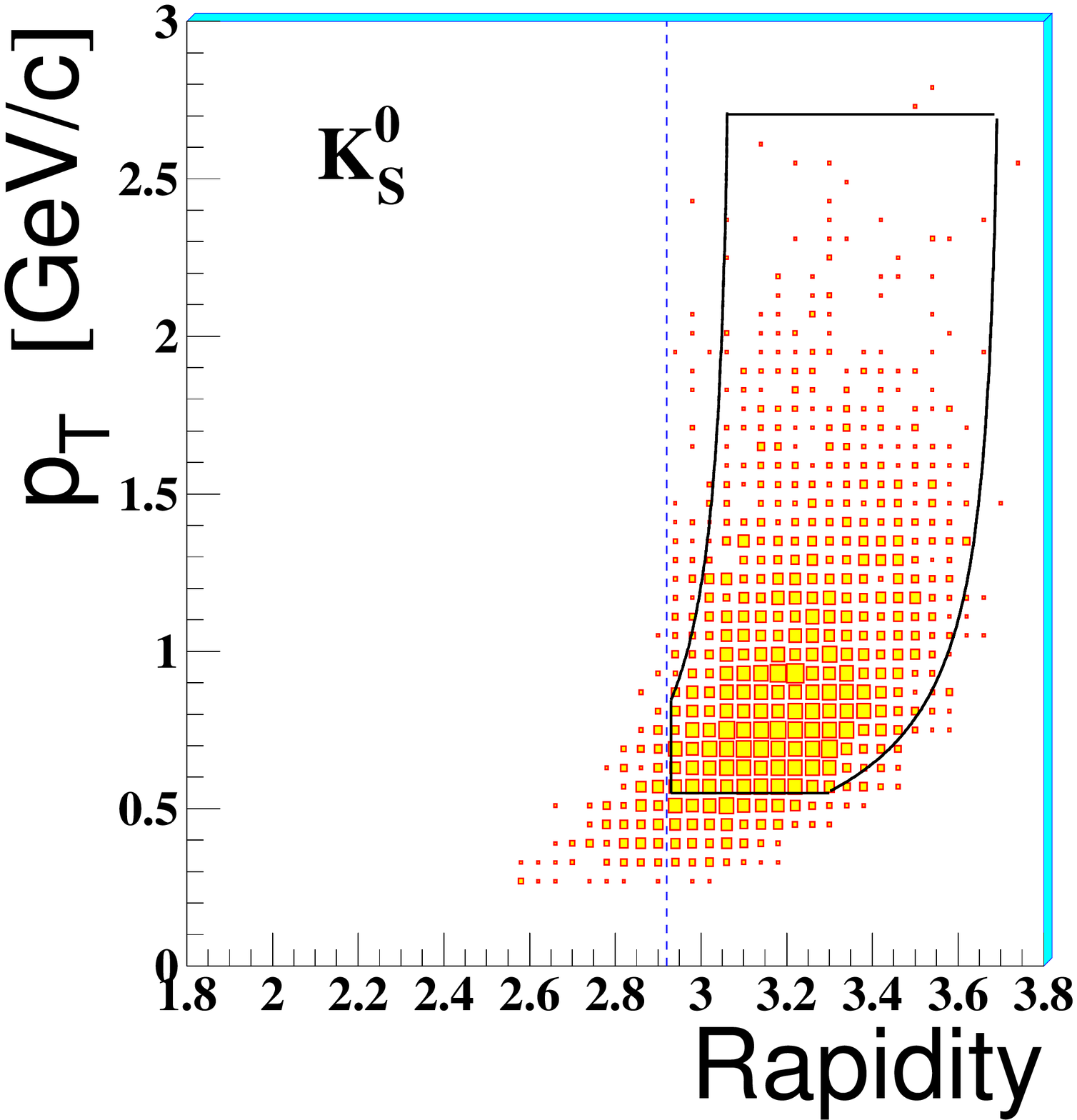}
\includegraphics{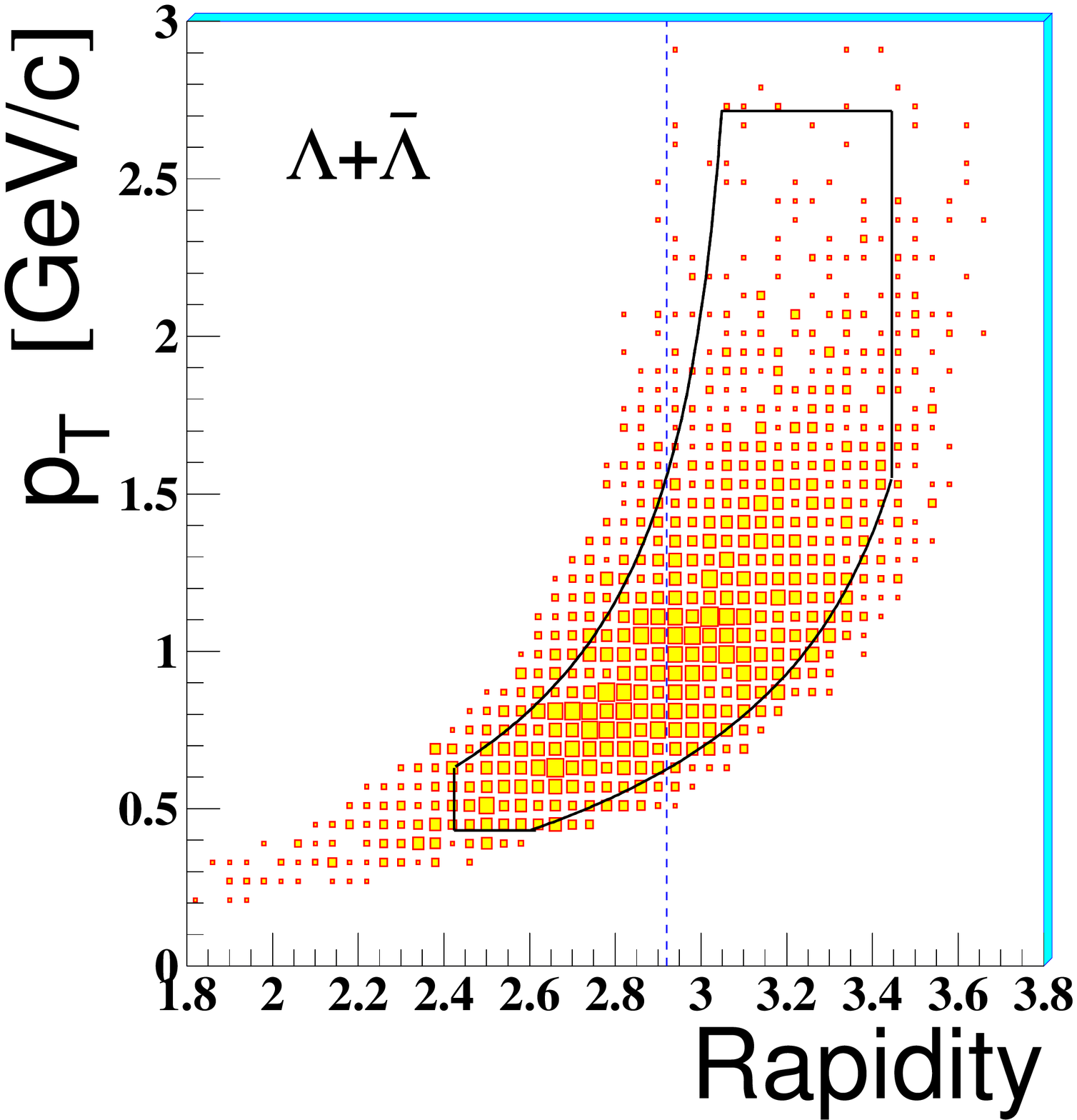}}\\
\resizebox{0.70\textwidth}{!}{%
\includegraphics{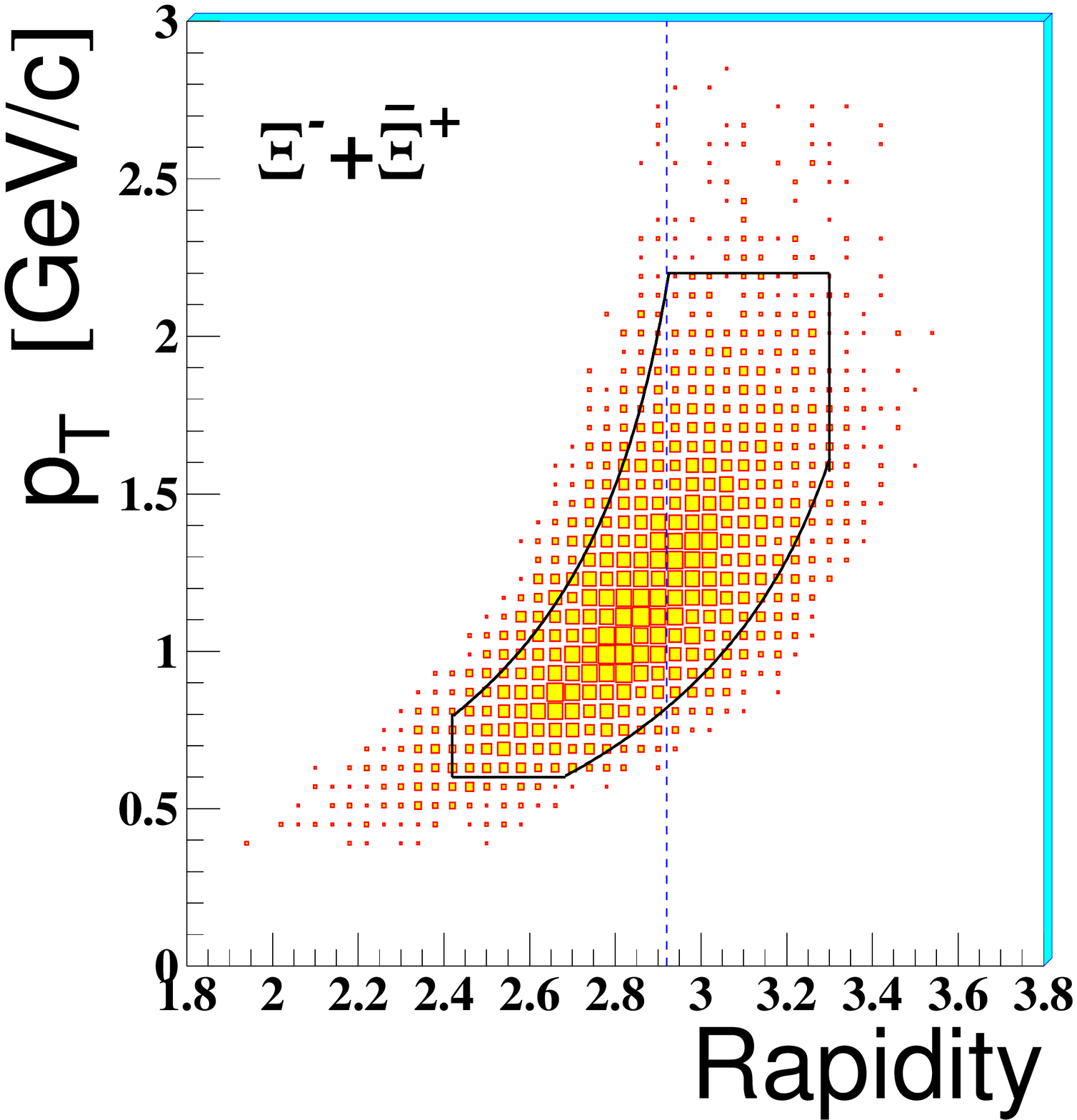}
\includegraphics{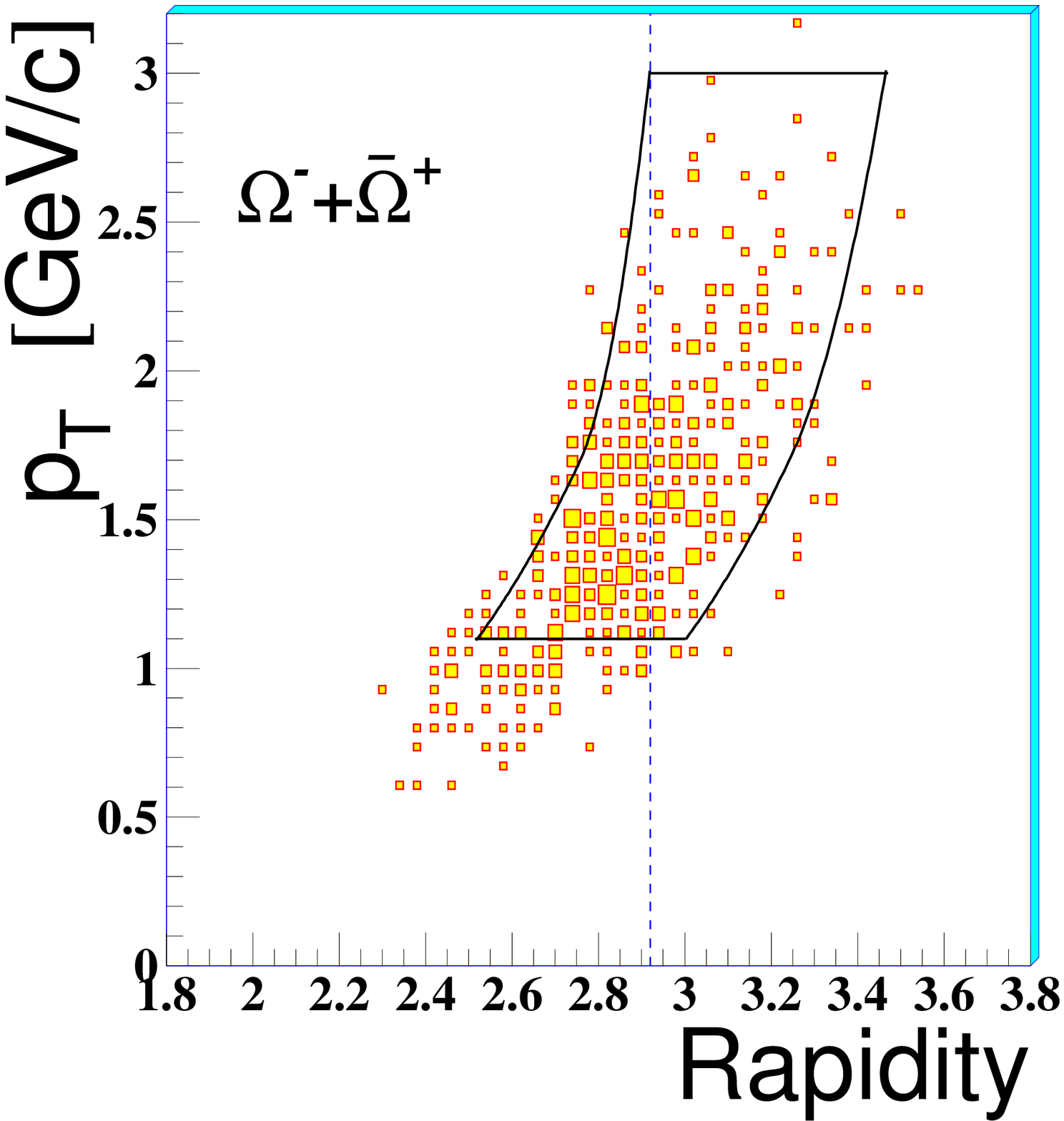}}
\caption{\rm The $y$--$p_{\tt T}$\  
         acceptance windows superimposed to the data samples.  
         Dashed lines show the position of mid-rapidity ($y_{cm}=2.92$).}
\label{fig:acceptance}
\end{figure}

For each particle species we define the fiducial  
acceptance window using a Monte Carlo simulation of the apparatus, 
in order to exclude the borders where the systematic errors  
are more difficult 
to evaluate.  
The 
acceptance regions are shown in figure~\ref{fig:acceptance}.  
For all hyperons, the acceptance window for particle and  
antiparticle 
are the same,   
due to the reflection symmetry of our apparatus with 
respect to the magnetic field direction. 
As a further check, the orientation of the magnetic field  
was periodically inverted.   

\begin{figure}[p]
\centering
\resizebox{0.85\textwidth}{!}{%
\includegraphics{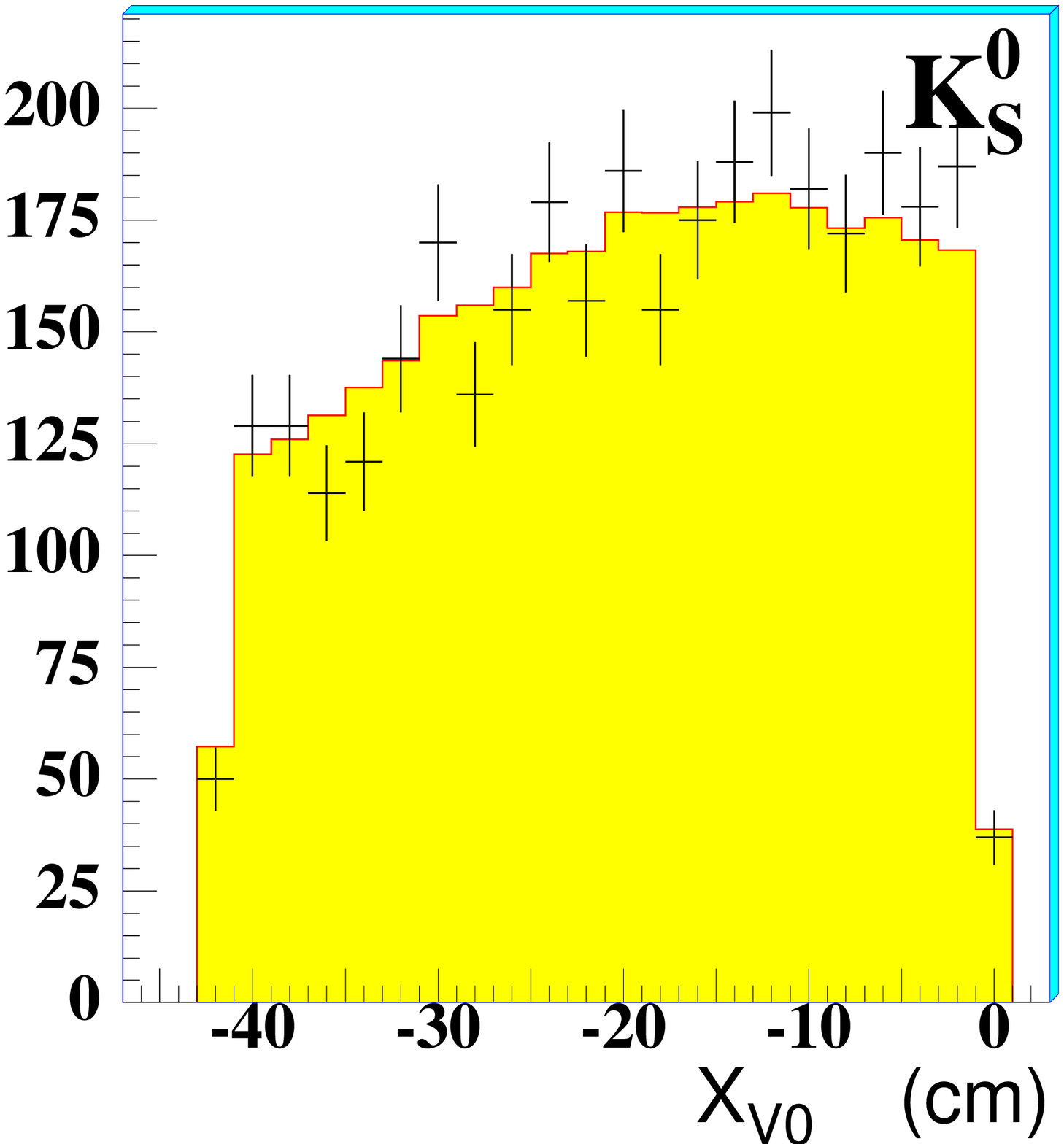}
\includegraphics{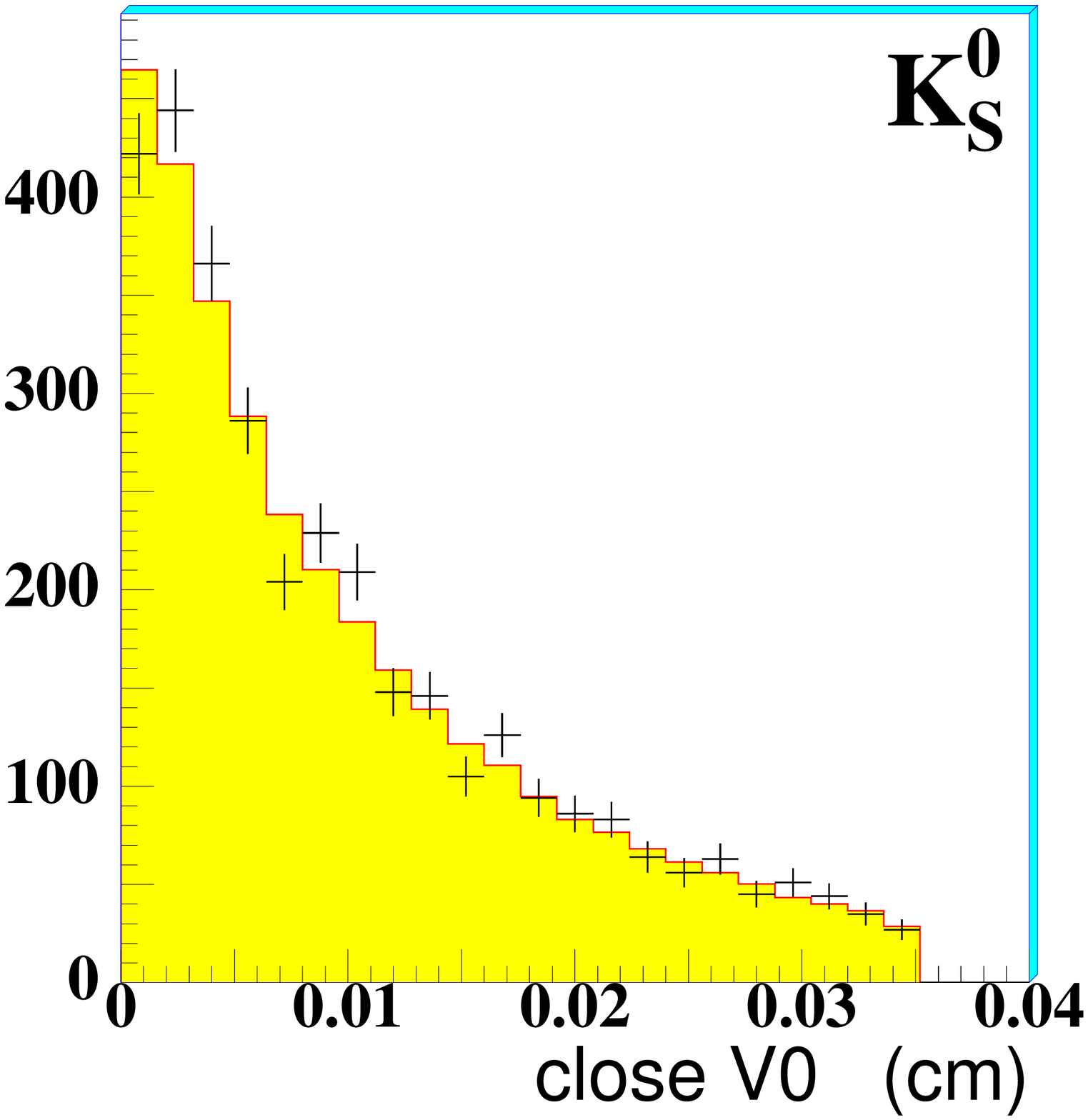}
\includegraphics{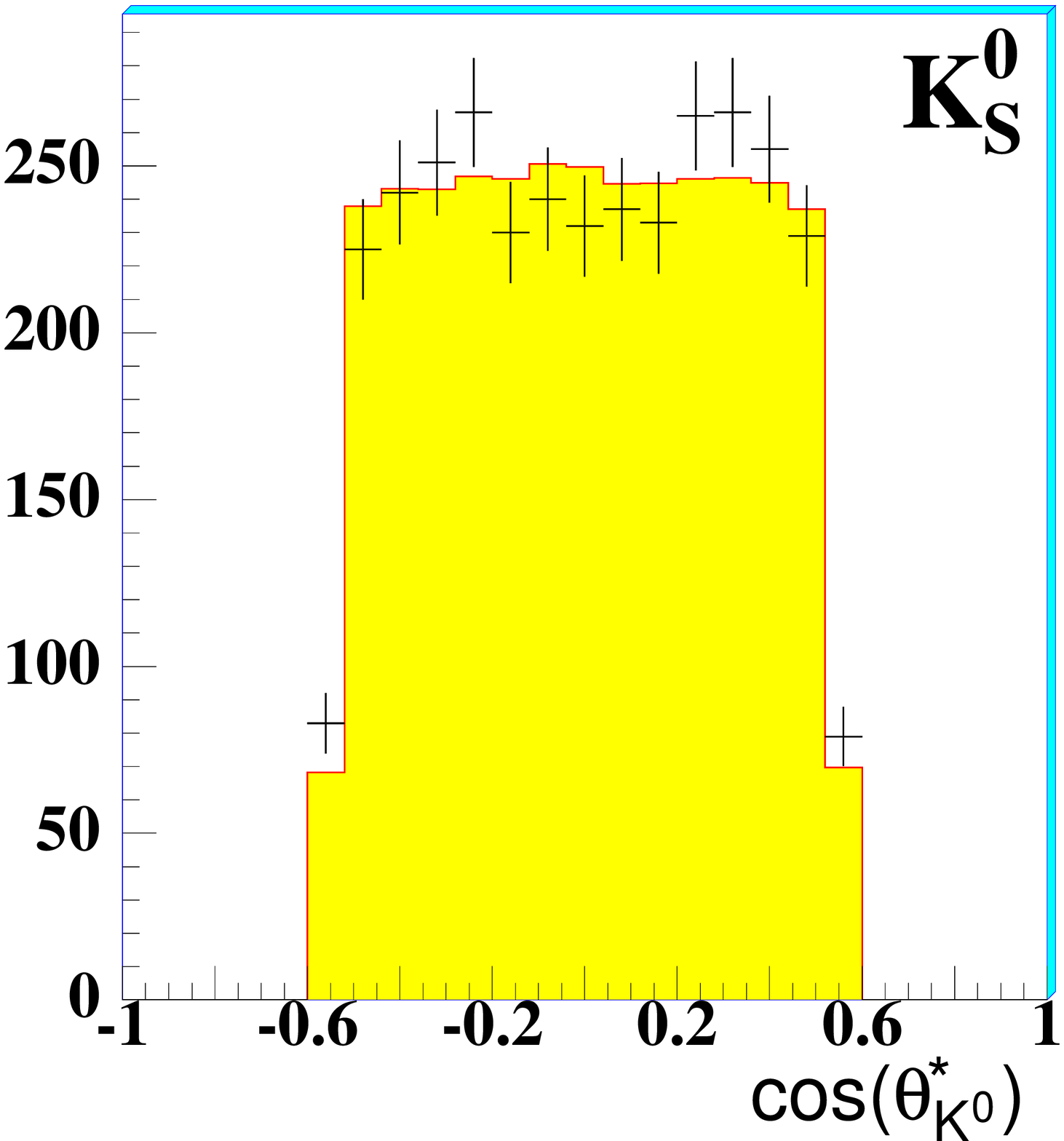}}\\
\resizebox{0.85\textwidth}{!}{%
\includegraphics{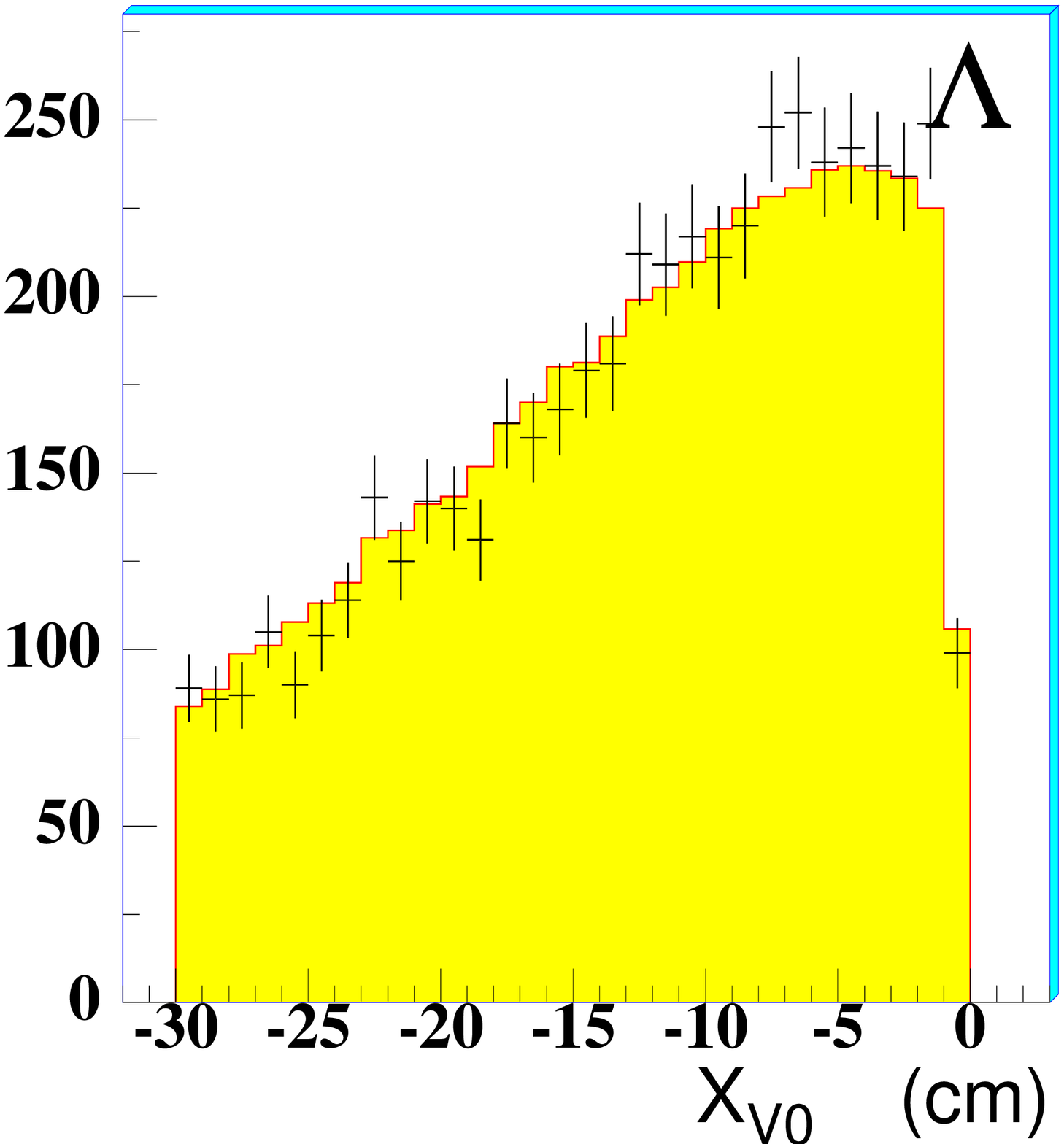}
\includegraphics{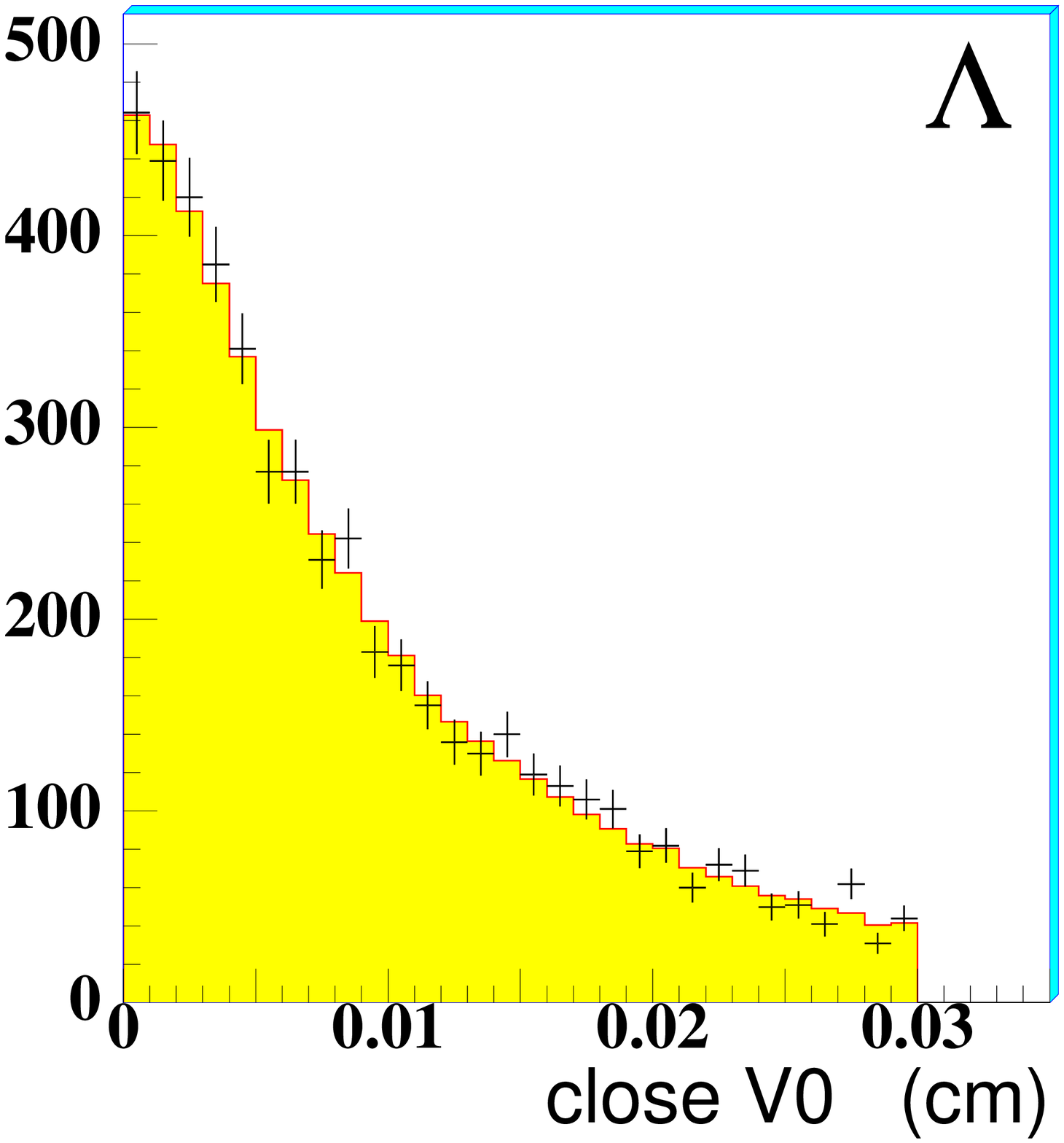}
\includegraphics{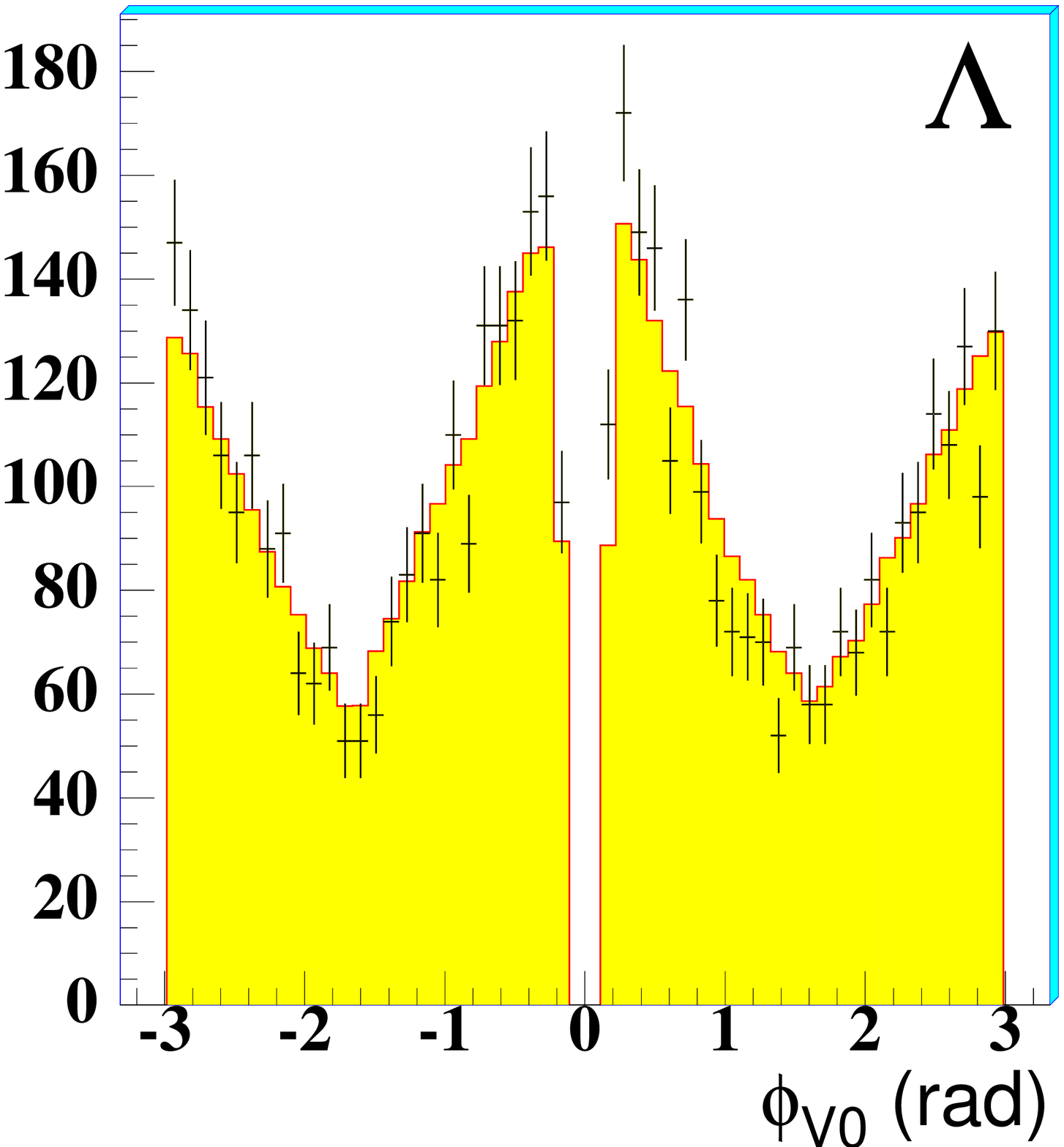}}\\
\resizebox{0.85\textwidth}{!}{%
\includegraphics{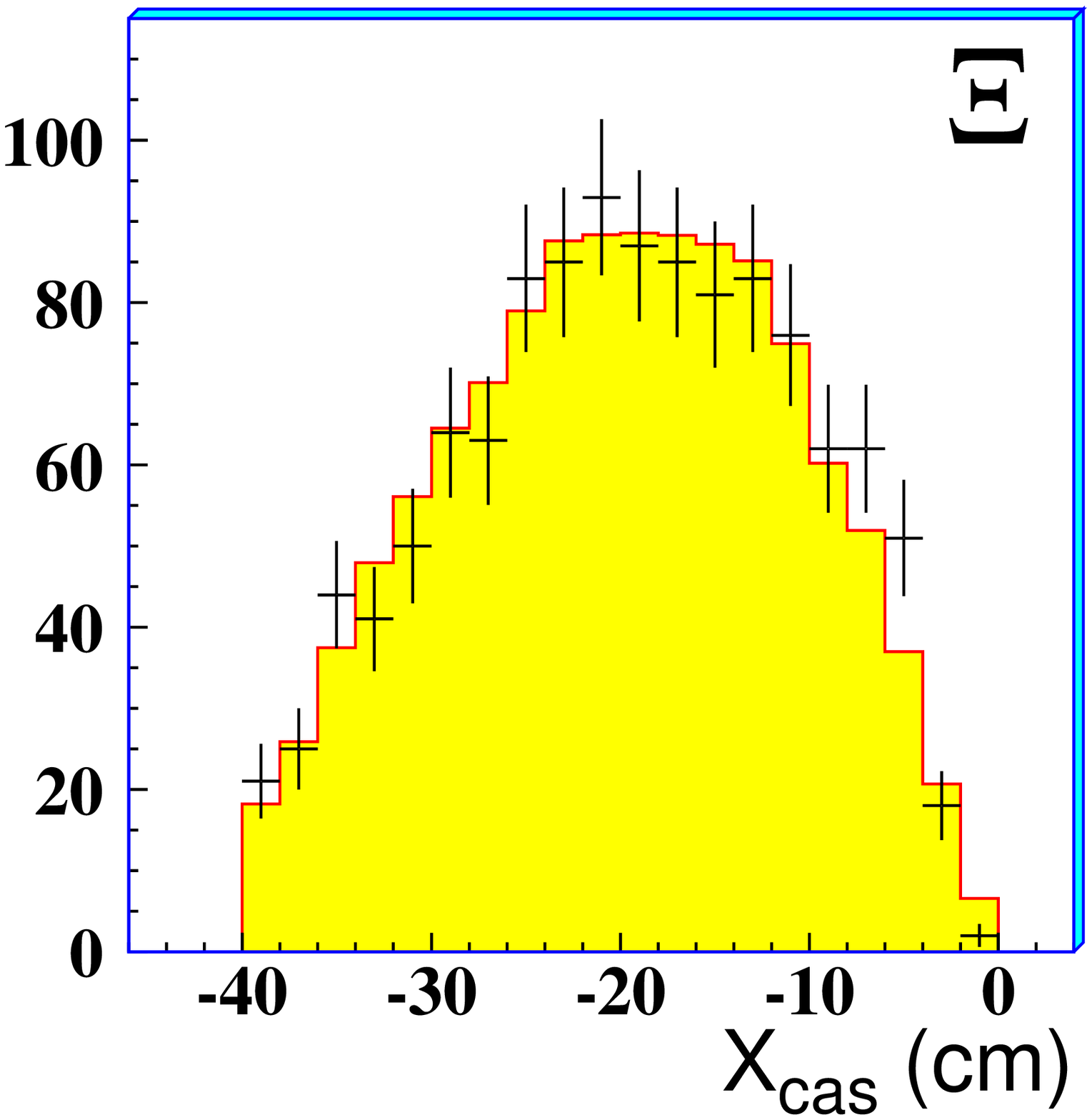}
\includegraphics{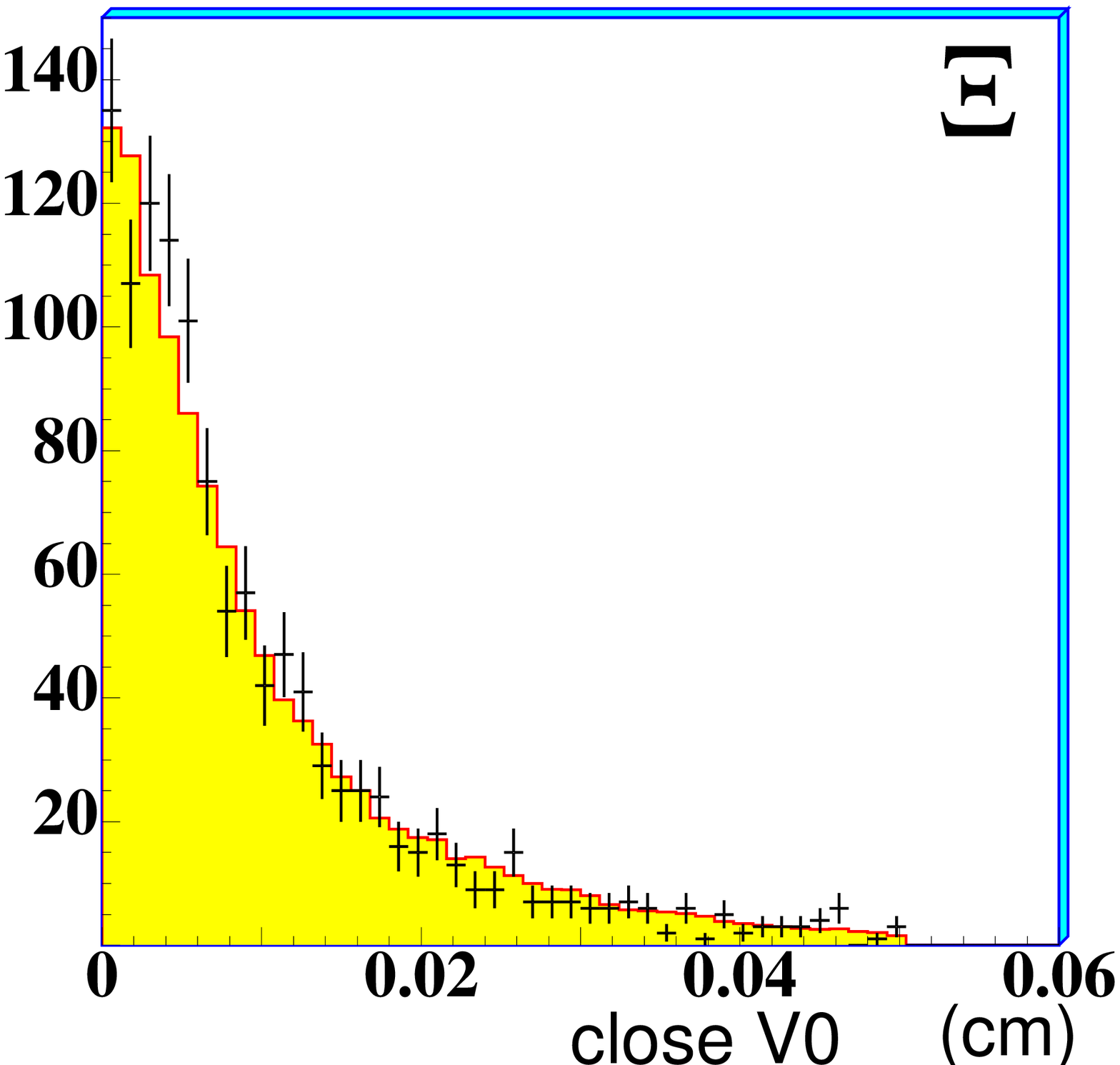}
\includegraphics{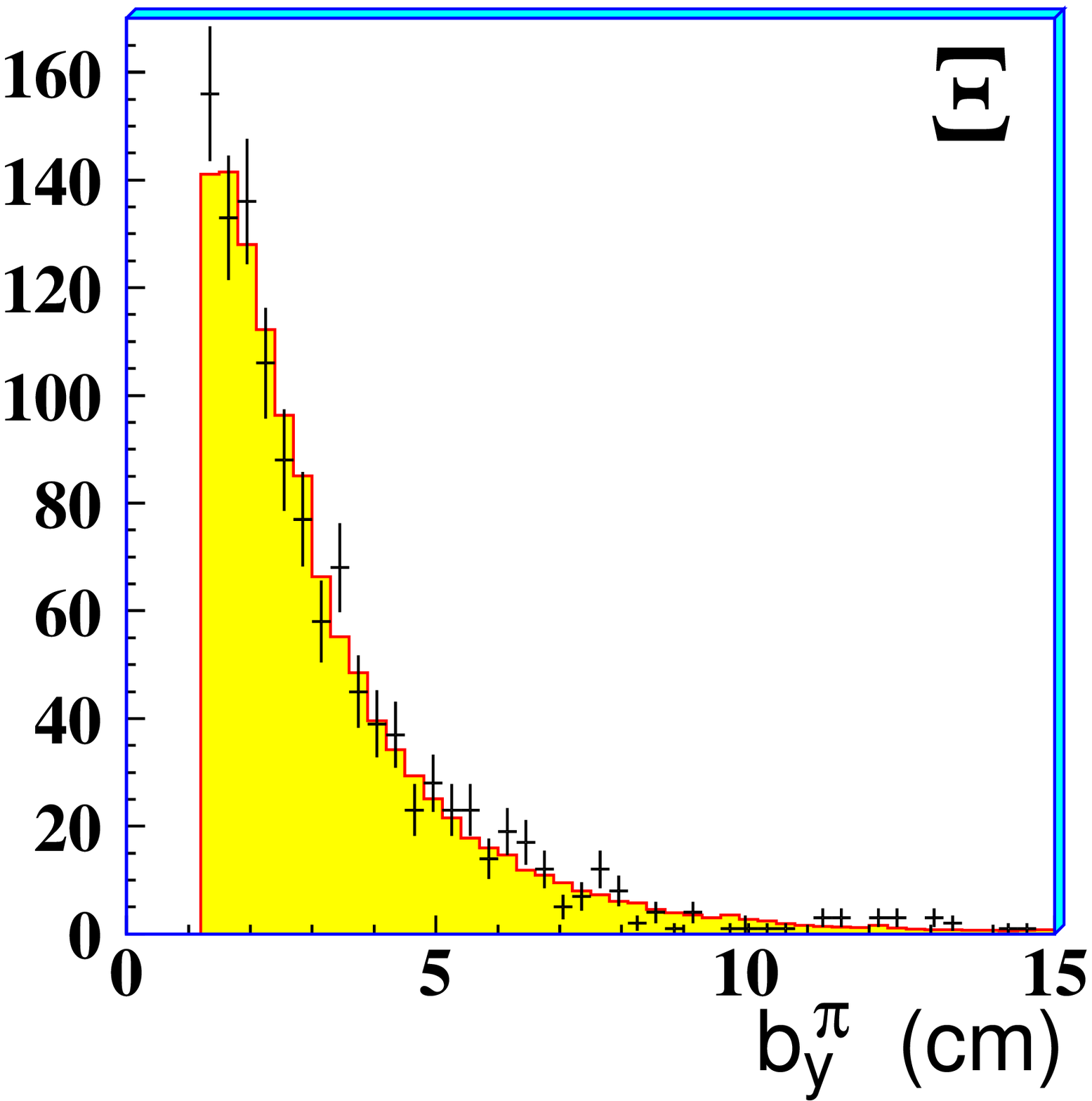}}\\
\resizebox{0.85\textwidth}{!}{%
\includegraphics{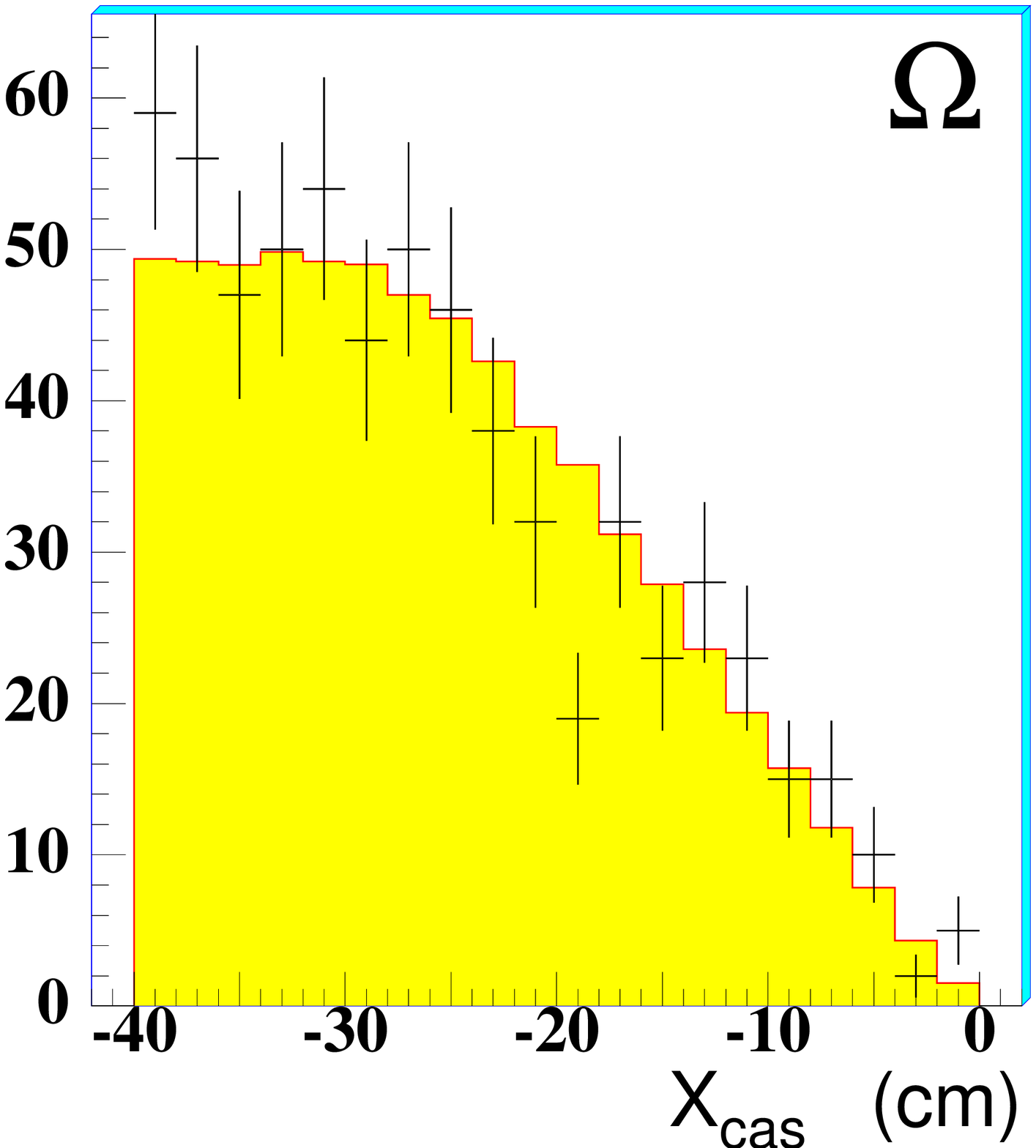}
\includegraphics{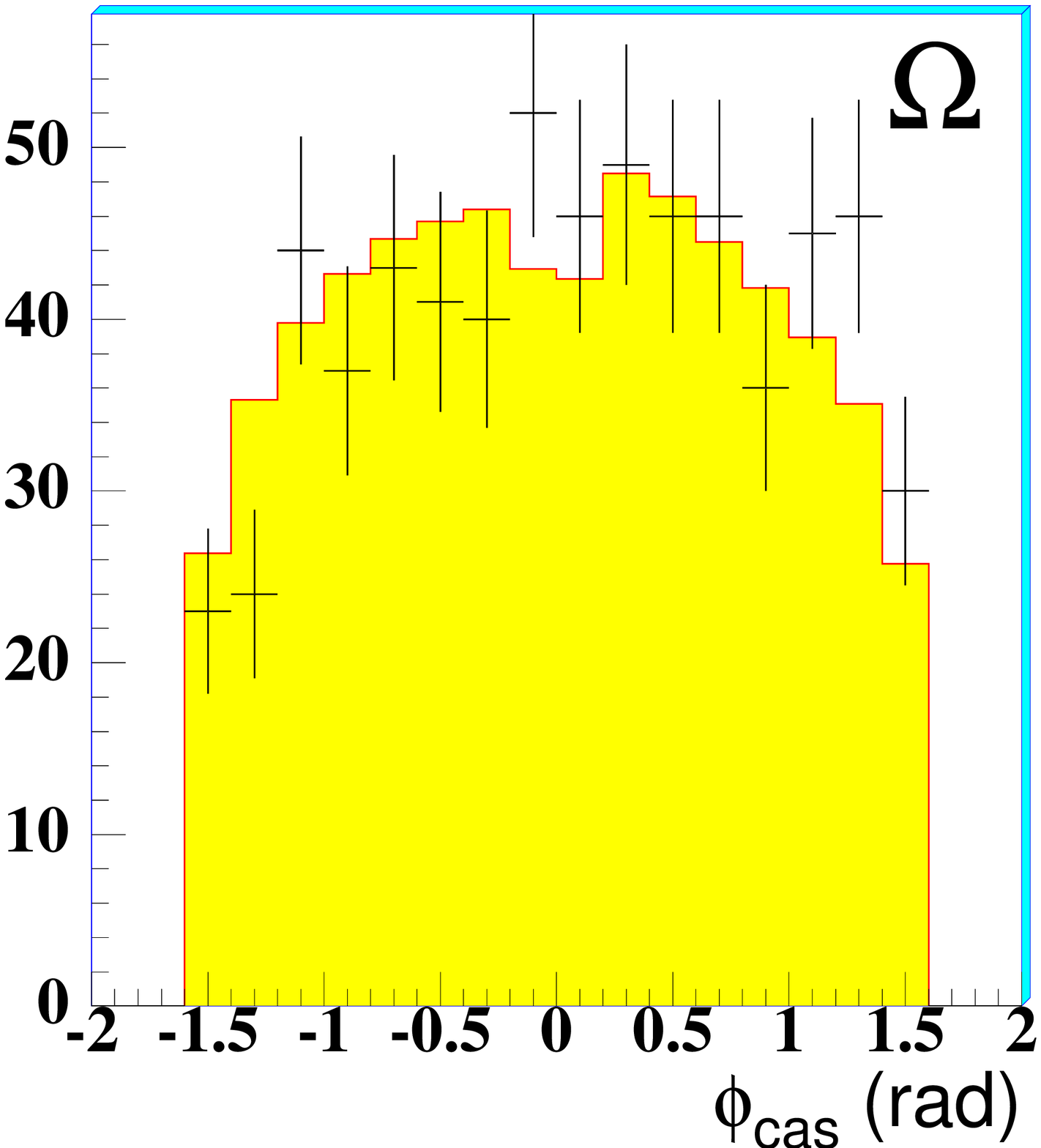}
\includegraphics{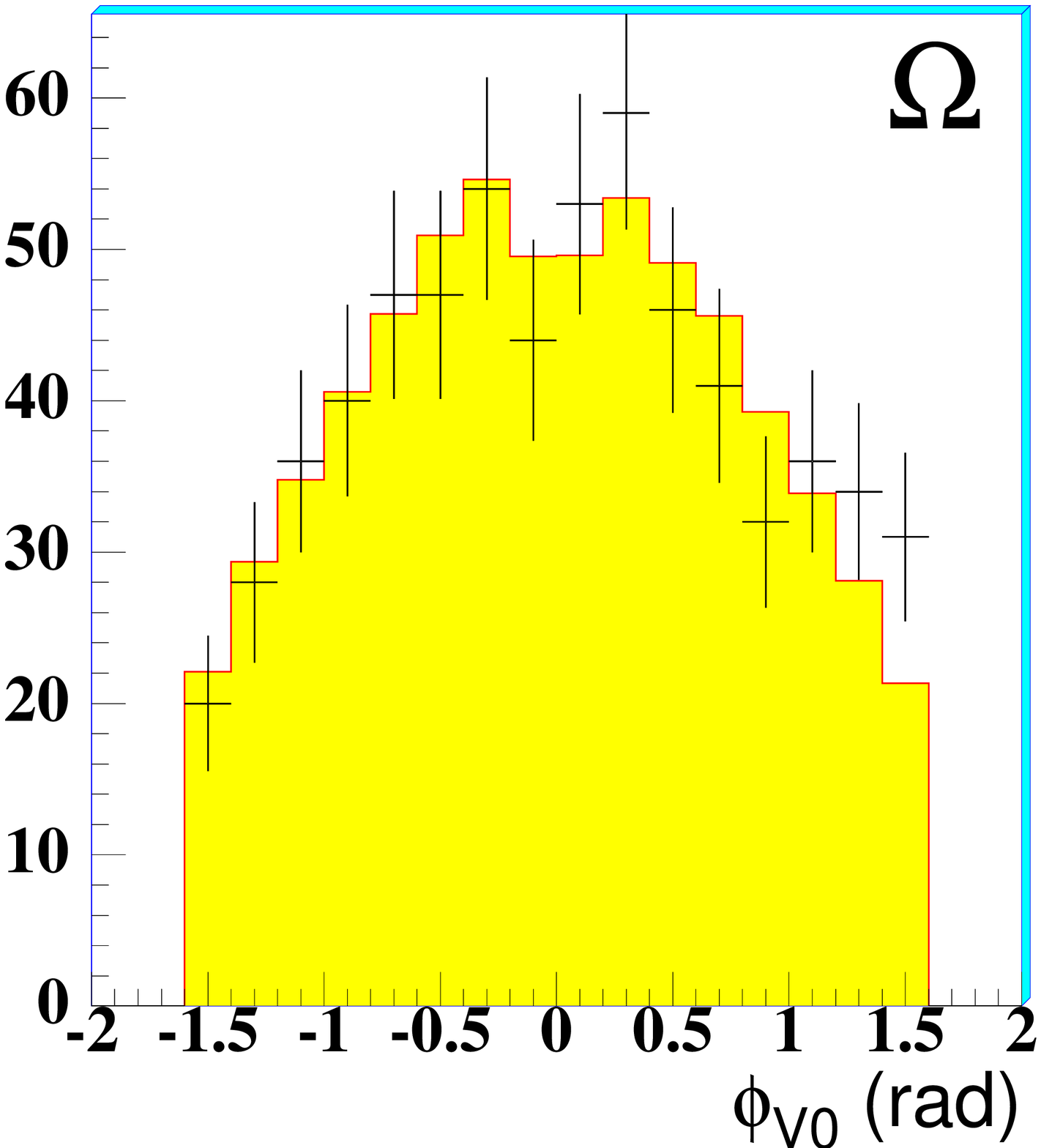}}\\
\caption{\rm Comparison between real (points with error bars)  
             and Monte Carlo (full lines) distributions. \;
{\bf $\bf 1^{st}$\ row}, \PKzS: the decay length of the \PKzS; the closest
                    distance in space ({\em close} parameter) 
		    between the extrapolated 
		    \Pgpp\ and \Pgpm\ tracks coming from the 
		    \PKzS\ decay; the cosine of $\theta^*$,  
		    the angle between the \Pgpm\ and the \PKzS\ lines of 
		    flight, in the \PKzS\ reference system. \;
{\bf $\bf 2^{nd}$\ row}, \PgL+\PagL: the decay length of the \PgL; 
                         the {\em close} parameter between the 
			 extrapolated \Pp\  and \Pgp;  
			 the azimuthal decay angle $\phi$. \;  
{\bf $\bf 3^{rd}$\ row}, \PgXm+\PagXp: the decay length of the 
                           $\Xi$; 
			   the {\em close} of the $\La$\ coming from the $\Xi$; 
                           the \Pgp\ impact parameter. \;  
{\bf $\bf 4^{th}$\ row}, \PgOm+\PagOp: the decay length of the $\Omega$; 
                           the azimuthal decay angle $\phi$\ in the $\Omega$\ 
			   decay; the azimuthal decay angle $\phi$\ in
			   the subsequent $\Lambda$\ decay.}
\label{fig:MonteCarloChecks}
\end{figure}

All data are corrected for geometrical acceptance and for detector and 
reconstruction inefficiencies on a 
event by event 
basis, with the 
following procedure: 
\begin{itemize}
\item for each reconstructed particle a sample of Monte Carlo particles  
      is generated with the 
      same measured rapidity $y$\ and transverse momentum $p_{\tt T}$, 
      random azimuthal angle 
      and the primary vertex position  
      generated randomly within the target according to the measured  
      beam profile;  
\item the Monte Carlo particles are traced through a  
      simulation of the apparatus
      \footnote{Silicon pixel detector efficiencies are accounted 
      for chip-by-chip. Each chip contains  2032 or 1006 
      sensitive cells (pixels) where the pixel size is, respectively,  
      $ 50 \times 500$\ $\mu$m$^2$\ or 
      $ 75 \times 500$\ $\mu$m$^2$~\cite{MANZ}.}
      based on GEANT~\cite{GEANT3}  
      allowing them to decay according to their
      life times and random internal decay angles.   
\item the hits of the decay tracks are merged with those of a real event --- with 
      a hit multiplicity close to that of the original event --- 
      in order to account for background tracks and electronic noise;  
\item the merged event is processed through the NA57 reconstruction and analysis chain;  
\item the weight to be associated with the real event is calculated as the ratio 
      of the number of Monte Carlo generated events to the number of 
      Monte Carlo events successfully reconstructed and retained by all 
      analysis selection criteria.   
\end{itemize} 
It has been checked that the final weights are not sensitive to the  
experimental smearing in $p_{\tt T}$\ and $y$. 
\newline
The simulation used for calculating the correction factors has been checked 
in detail~\cite{Kristin} by comparing real and Monte Carlo distributions 
for several parameters, and they show good agreement.  
As an example, we show in figure~\ref{fig:MonteCarloChecks} 
three such distributions for each particle species.  

The systematic errors on the inverse slopes have been estimated to be 
about 10\% for all the strange particles.  

The weighting method described above is 
CPU intensive; therefore, while each of the reconstructed $\Xi$s and $\Omega$s 
have been individually weighted,  
for the much more abundant \PKzS, \PgL\ and \PagL\ 
samples we only weighted a small fraction of the total 
sample in order to reach a statistical accuracy better than the limits 
imposed by the systematic error.  
Table~\ref{tab:sample} shows the strange particle samples collected by NA57 and 
those used for the present analysis.  
\begin{table}[h]
\caption{Statistics of the strange particle samples 
         used in this analysis (individually {\em weighted}) and  
         {\em collected} by NA57 in Pb-Pb collisions at 158 $A$\ GeV/$c$.  
\label{tab:sample}}
\begin{center}
\begin{tabular}{|c|ccccccc|}
\hline
 \hline
    Particle  & \PKzS & \PgL & \PagL & \PgXm & \PagXp & \PgOm & \PagOp \\
{\em weighted}& 3340  & 2350 & 2718  & 5858  & 1522   &  432  &  192   \\
{\em collected}& $ \times 400$
                   & $ \times 400$  
          & $\times50$ & $\times1$ & $\times1$ & $\times1$ & $\times1$ \\
\hline
\end{tabular}
\end{center}
\end{table}
%

\begin{figure}[hbt]
\centering
\resizebox{0.40\textwidth}{!}{%
\includegraphics{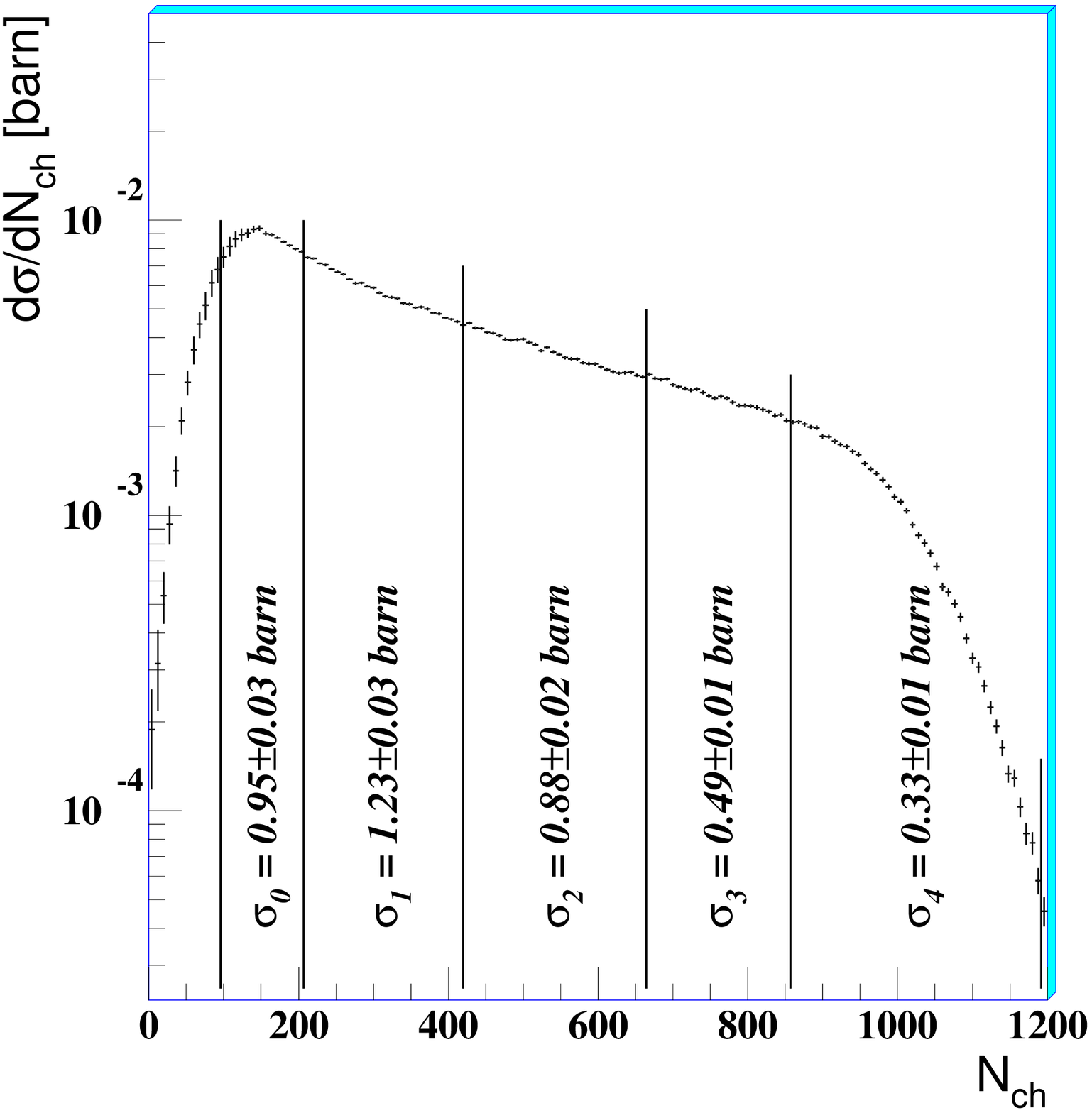}}
\caption{\rm  Charged particle multiplicity distribution   
              and limits of the five centrality classes.  
	      For each class the corresponding   
	      cross-section is indicated.}  
\label{fig:multiplicity}
\end{figure}

The charged particle multiplicity distribution, 
which is shown in figure~\ref{fig:multiplicity},  
has been divided into five centrality classes (0,1,2,3,4), class 0 
being the most peripheral and class 4 the most central. The drop at  
low multiplicities is due to the centrality selection applied at 
the trigger level.  
The fractions of the inelastic cross-section for the five classes, 
calculated assuming a total cross section of 
$7.26$\  barn,   
are given in table~\ref{tab:centrality}.  
\noindent
\begin{table}[h]
\caption{Centrality ranges for the five  classes defined in figure~\ref{fig:multiplicity}.
\label{tab:centrality}}
\begin{center}
\begin{tabular}{|c|ccccc|}
\hline
 Class &   $0$   &   $1$   &   $2$   &  $3$   &   $4$ \\
 $\sigma/\sigma_{inel}$\ \; (\%)   & 53--40 & 40--23 & 23--11 & 11--4.5 & 4.5--0 \\
\hline
\end{tabular}
\end{center}
\end{table}
\noindent
\newline
The procedure for the measurement of the multiplicity distribution and  
the determination of the collision centrality for each class  
is described in reference~\cite{Multiplicity}. 
\section{Inverse slopes of the transverse mass spectra}
The double-differential $(y,m_{\tt T})$\ distribution for each particle  
species has been parametrized using the expression  
\begin{equation}
\label{eq:expo}
\frac{d^2N}{m_{\tt T}\,dm_{\tt T} dy}=f(y) \hspace{1mm} \exp\left(-\frac{m_{\tt T}}{T_{app}}\right)
\end{equation}
assuming the rapidity distribution to be flat within our acceptance region
($f(y)={\rm const}$).  
The inverse slope parameter $T_{app}$\ (``apparent temperature'') 
has been extracted by means of a maximum likelihood fit of equation~\ref{eq:expo} 
to the data. As discussed in the next section, this apparent temperature 
is interpreted as due to the  
thermal motion coupled with a collective transverse flow  
of the fireball components assumed to be in thermal equilibrium.  
\newline
In the fits of the $\PgOm$\ and $\PagOp$\ particle spectra,  
the lower limit for the accepted transverse momentum was chosen to be 1.4 GeV/$c$. 
This choice allows us to exclude a possible instrumental bias at low $p_{\tt T}$\ 
for $\Omega$.  
The values of the transverse expansion velocity and of the freeze-out 
temperature obtained within the blast-wave model (see section 5) 
do not depend significantly on the choice of the lower limit.   

\begin{figure}[hbt]
\centering
\resizebox{0.74\textwidth}{!}{%
\includegraphics{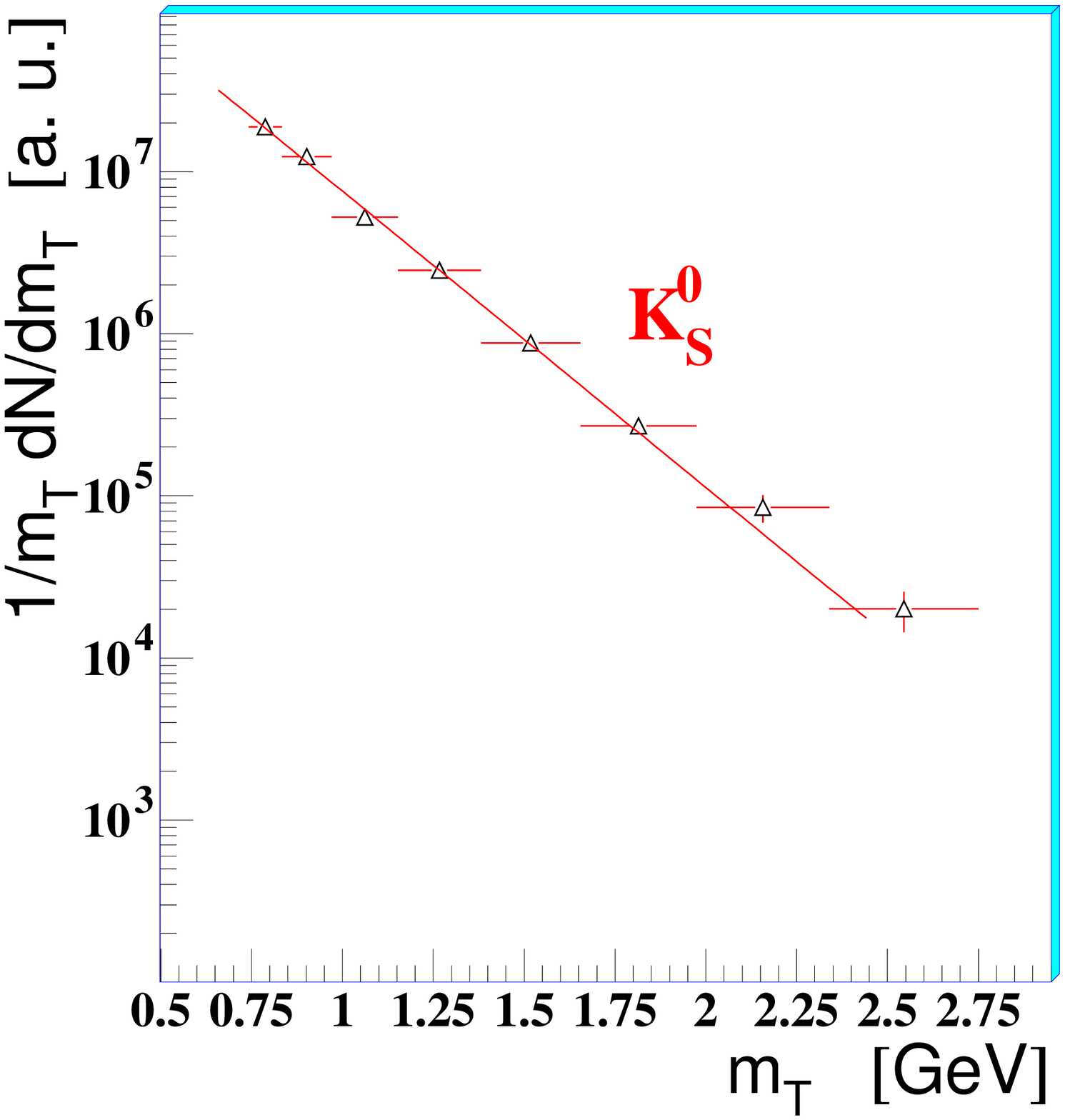}
\includegraphics{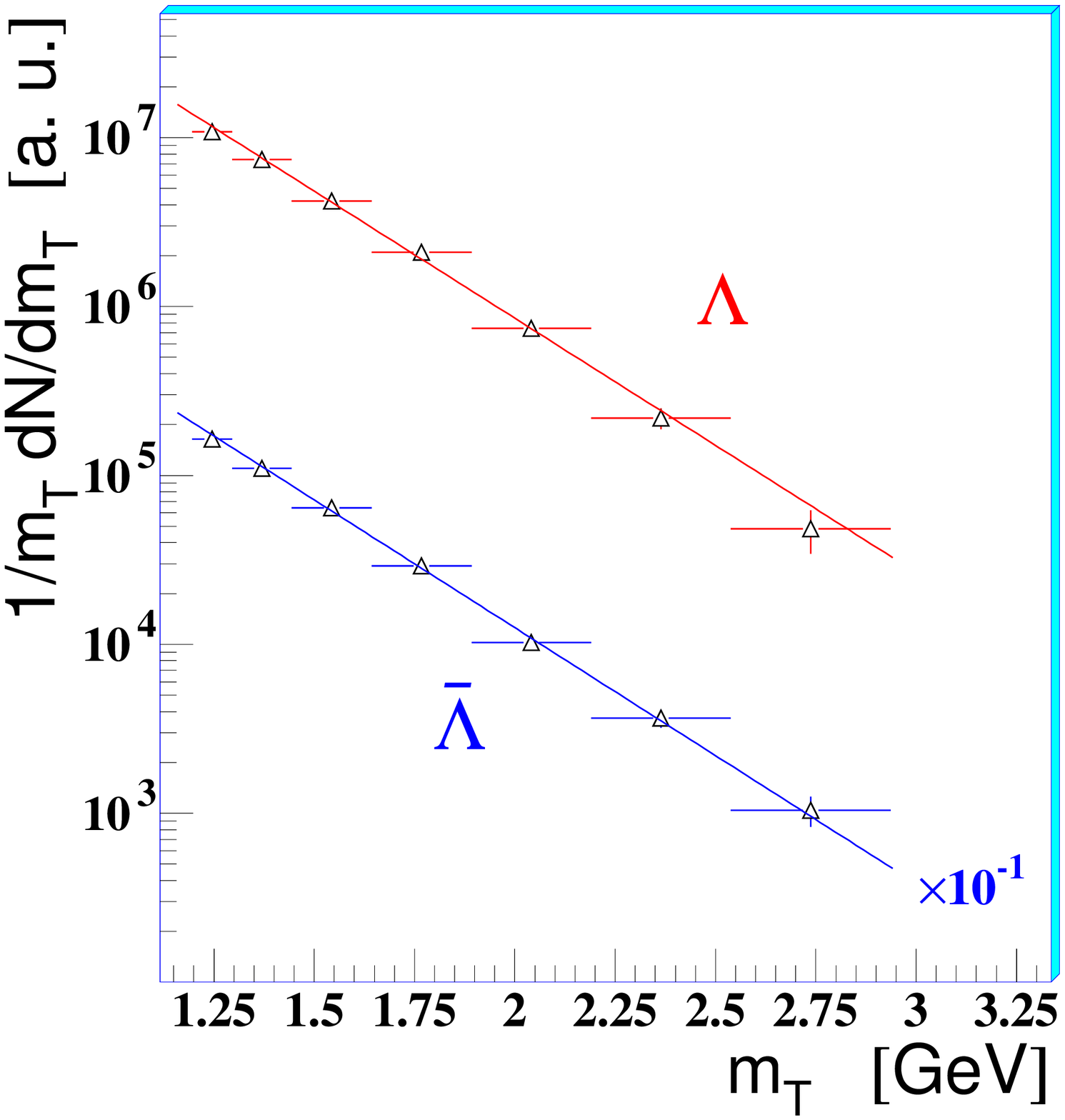}}\\
\resizebox{0.74\textwidth}{!}{%
\includegraphics{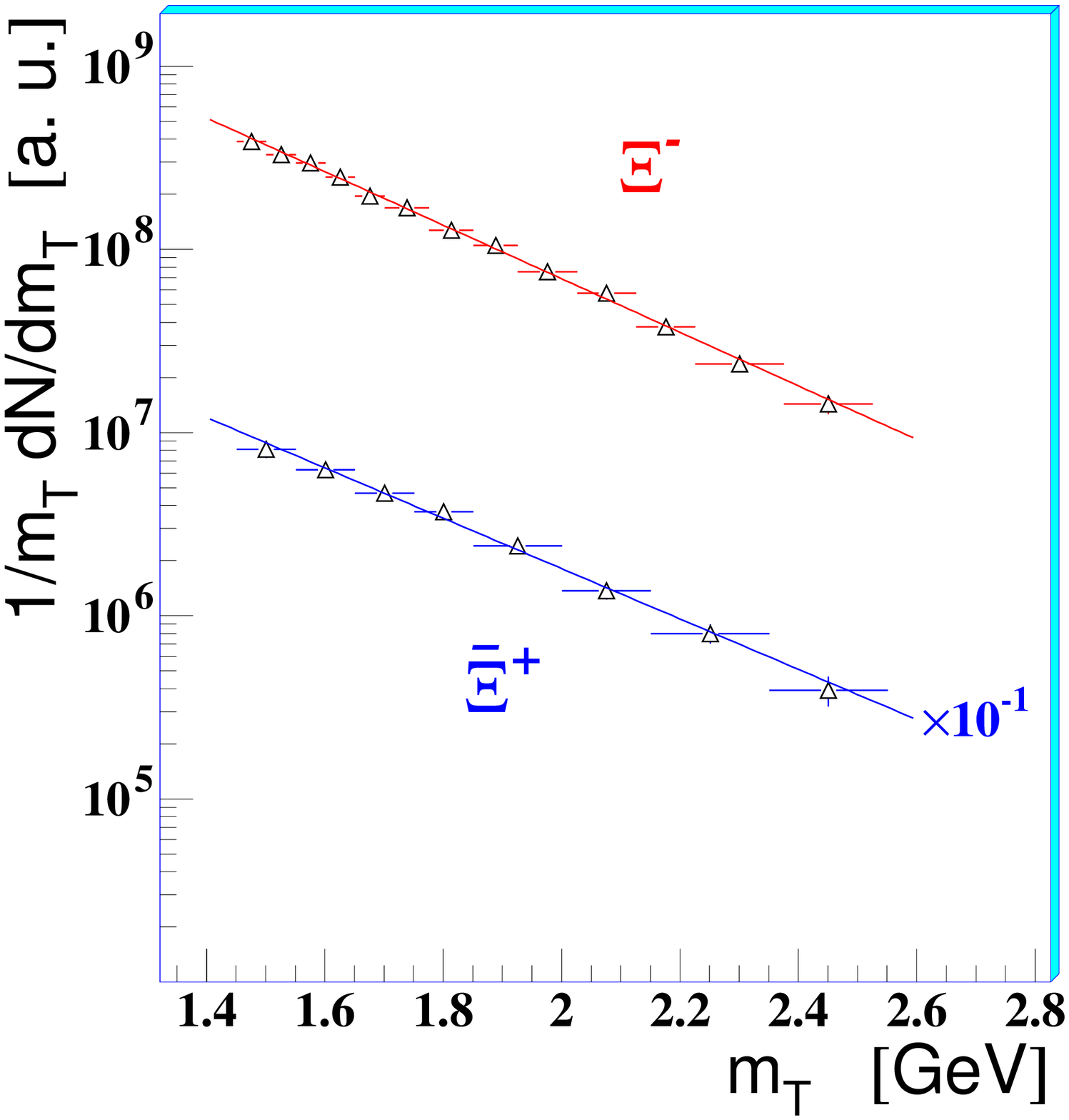}
\includegraphics{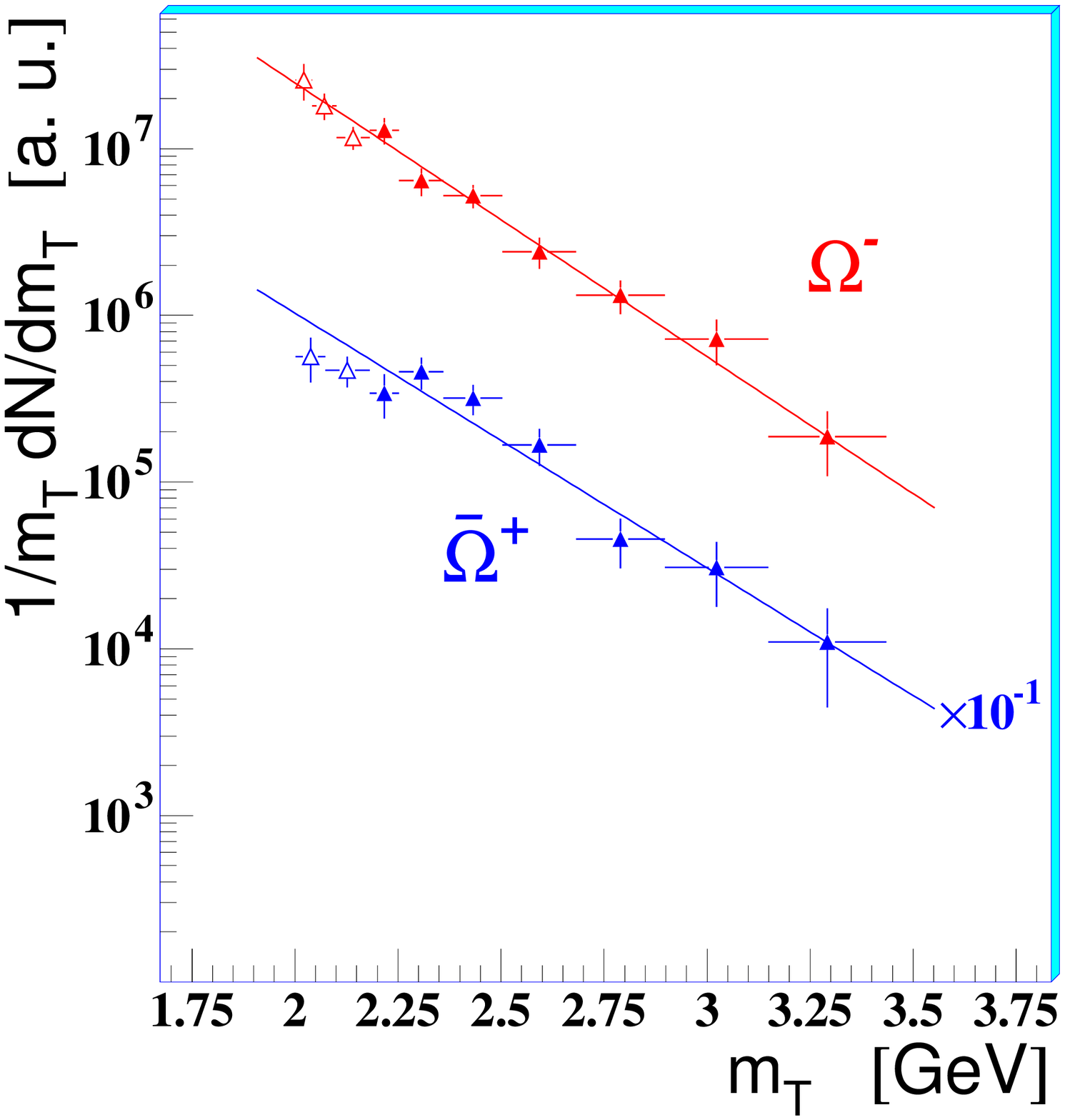}}
\caption{\rm  Transverse mass spectra of strange particles 
 for the 53\% most central Pb-Pb cross-section.   
 The superimposed exponential functions have inverse slopes equal to the  
 $T_{app}$\ values obtained from the maximum likelihood best fits. 
 In the \PgOm\ and \PagOp\
 spectra, the open points have not been used in the best fit calculation.}  
\label{fig:all_spectra}
\end{figure}

The $1/m_{\tt T} \, dN/dm_{\tt T} $\ distributions for strange particles 
measured for the full centrality  
range under consideration are shown in figure~\ref{fig:all_spectra}.   
%
The shapes of all spectra are well described by exponential functions.   
In the next section, we exploit the deviations from   
the exponential in order to disentangle the transverse flow 
from the thermal motion.  
The inverse slope parameters $T_{app}$\ of the transverse mass spectra are 
given in table~\ref{tab:InvSlopes}.  They are in agreement within the 
errors with those measured over a smaller centrality range 
(about the 40\% most central inelastic Pb-Pb cross-section)   
by the WA97 experiment~\cite{MtWA97}.   
\begin{table}[htb]
\caption{Inverse slope parameter $T_{app}$\ (MeV) of the strange particles
in the full centrality range ({\bf 0-4}). The first error is statistical, the 
second one systematic. 
\label{tab:InvSlopes}}
\begin{center}
\begin{tabular}{|c|c|c|}
\hline
           {\bf $K_S^0$}  &   {\bf $\La$}  &  {\bf $\Al$}   \\ \hline
	   $237\pm4\pm24$ & $289\pm7\pm29$ & $287\pm6\pm29$ \\ \hline	 
\end{tabular}

\begin{tabular}{|c|c|c|c|} \hline
     {\bf $\Xi^-$}  & {\bf $\overline\Xi^+$} & {\bf $\Omega^-$}  
                    & {\bf $\overline\Omega^+$} \\ \hline
     $297\pm5\pm30$ & $316\pm11\pm30$ & $264\pm19\pm27$   
                    & $284\pm28\pm27$ \\ \hline
\end{tabular}
\end{center}
\end{table}

\begin{figure}[hbt]
\centering
\resizebox{0.60\textwidth}{!}{%
\includegraphics{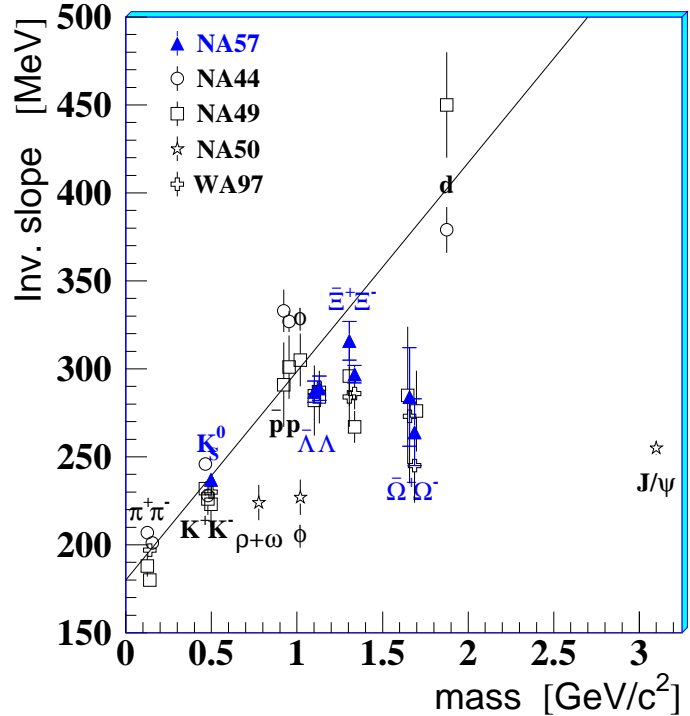}}
\caption{\rm Inverse slopes as a function of the particle rest mass  
 in Pb-Pb interactions at 158 $A$\ GeV/$c$\ (see text for references). 
 Only statistical errors are shown. 
 The line through $\pi$, K and proton is shown   
 to guide the eye.}  
\label{fig:Inv_mass}
\end{figure}

A compilation of data\footnote{ 
NA49 results are taken from the following references: \Pgpm, \PKm\ and \PKp\  
from~\cite{ref:138}; \PKzS\ from~\cite{ref:139}; \Pgpp, \Pp\ and \Pap\  
from~\cite{ref:140}; $\phi$\ from~\cite{ref:141}; \PgL\ and \PagL\  
from~\cite{ref:142}; \PgXm\ and \PagXp\ from~\cite{ref:143};  
\PgOm\ and \PagOp\ from~\cite{ref:144}; 
deuteron from~\cite{ref:145}. 
NA44 results are collected for \Pgpm, \Pgpp, \PKm, \PKp, \Pp, \Pap\ 
from reference~\cite{ref:146} 
and for the deuteron from reference~\cite{ref:147}.   
The point at $m \simeq 3.1$\ GeV/$c^2$\ corresponds to the NA50 determination 
of a thermal spectrum  for the $J/\psi$\ charmonium state~\cite{ref:47}.   
The $J/\psi$\ spectra are fitted 
using a modified Bessel function, namely $1/T\cdot m_{\tt t}^2 \cdot K_1(m_{\tt t}/T)$; this 
yields an inverse slope $T=255\pm4$\ MeV (the value showed 
in figure~\ref{fig:Inv_mass})  for the most central collisions, 
which is smaller by about 5--10 MeV than the one obtained 
with the exponential function~\cite{ref:47}.  
Finally the NA50 inverse slopes
for the $\rho+\omega$\ and $\phi$\ vector-mesons 
are taken from reference~\cite{NA50_rho_w_phi}.
}
on the $m_{\tt T}$\  
inverse slopes as a function of the particle mass in Pb-Pb interactions at 
158 $A$\ GeV/$c$\ is shown in figure~\ref{fig:Inv_mass}.   
\newline
As already found by the WA97 experiment, the inverse slopes of the $\Omega$\ hyperon  
and of the $\Xi$\  
fall below the line drawn through the $\pi$, K and proton points.  
This observation has been interpreted as due to an early freeze-out of    
multi-strange hadrons~\cite{Hecke}.  
%
\subsection{Centrality dependence}
We have fitted the weighted 
data by equation~\ref{eq:expo}  
separately for each of the five multiplicity classes indicated in  
figure~\ref{fig:multiplicity}.  
\begin{figure}[t]
\centering
\resizebox{0.94\textwidth}{!}{%
\includegraphics{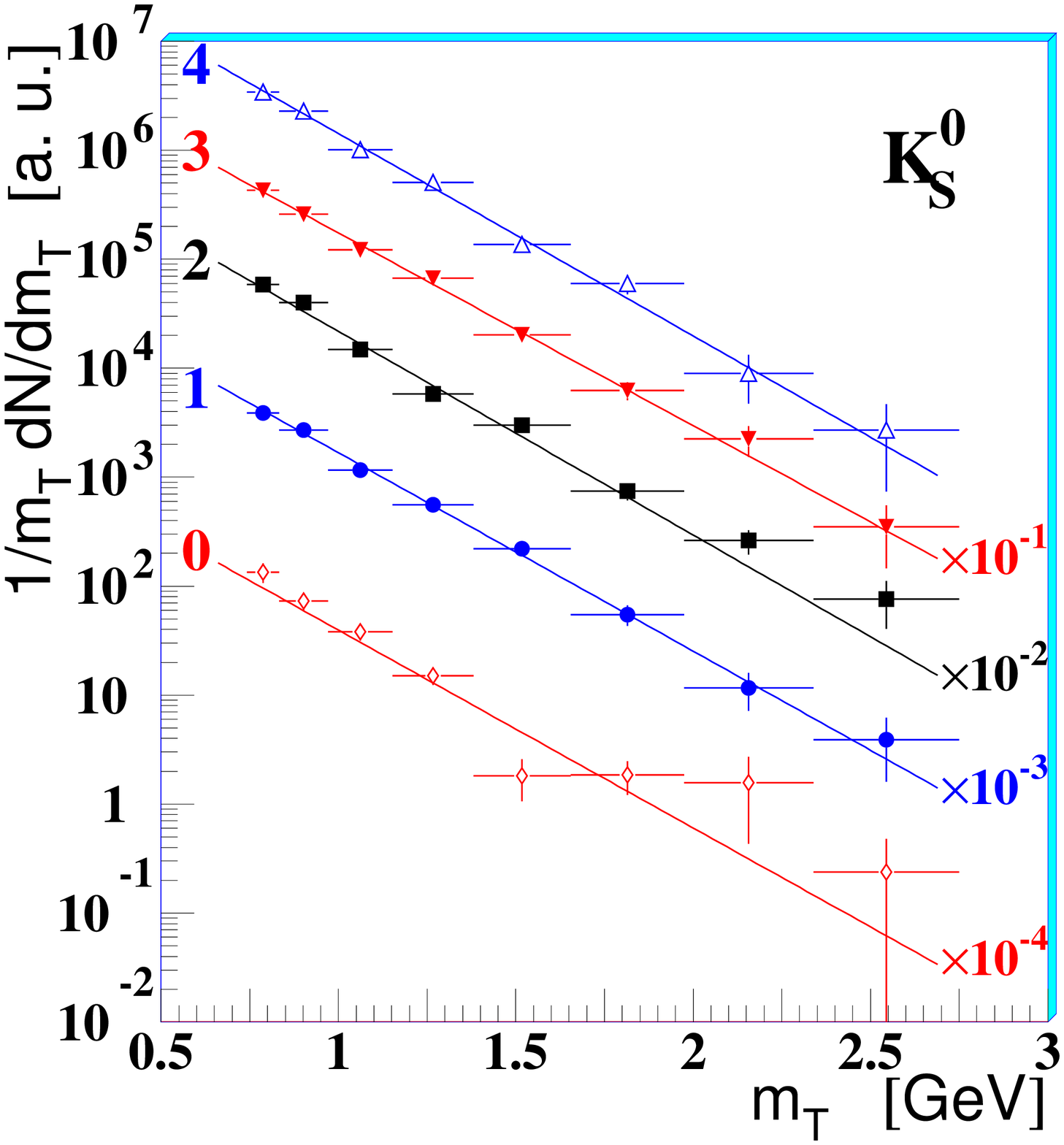}
\includegraphics{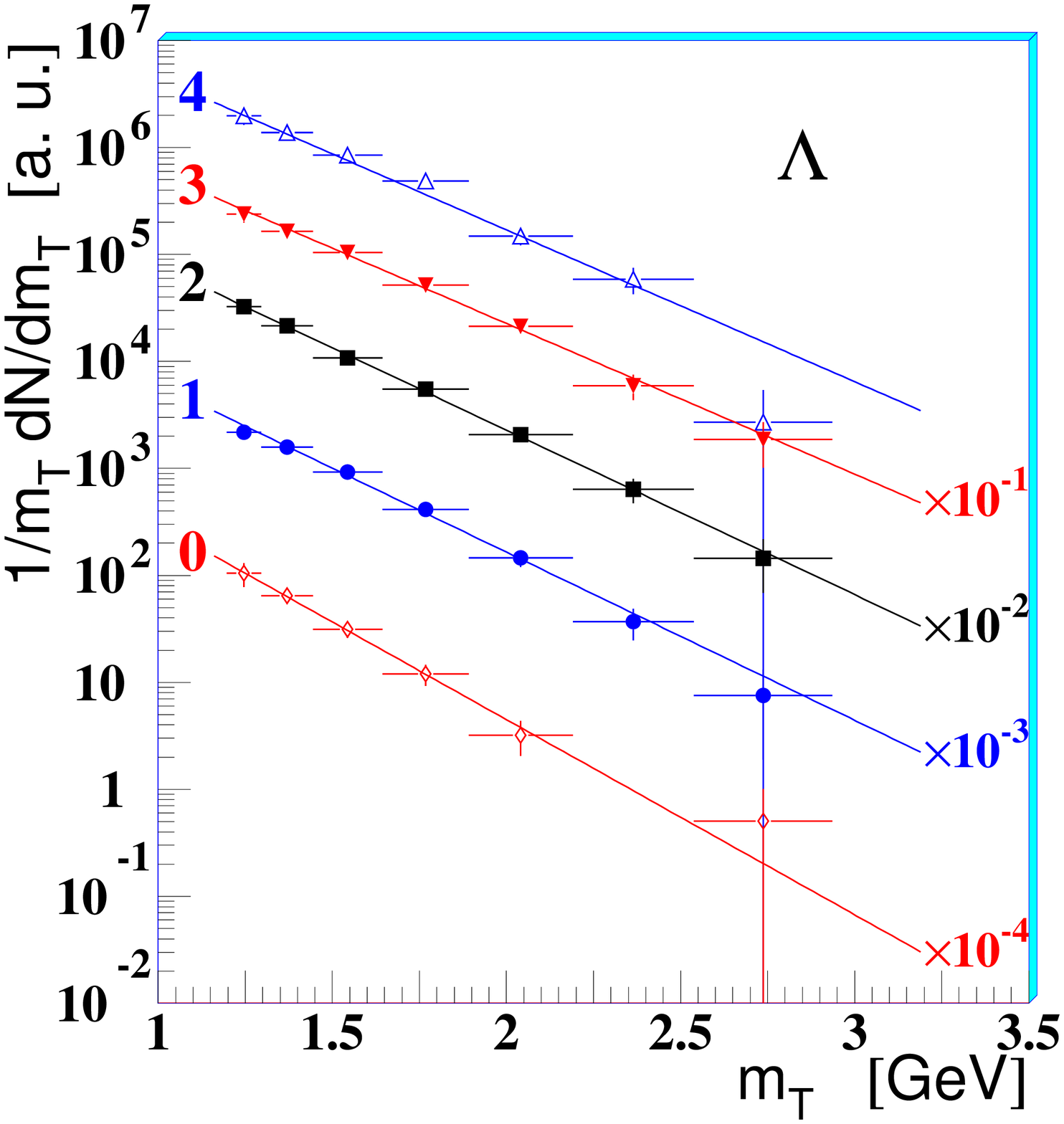}
\includegraphics{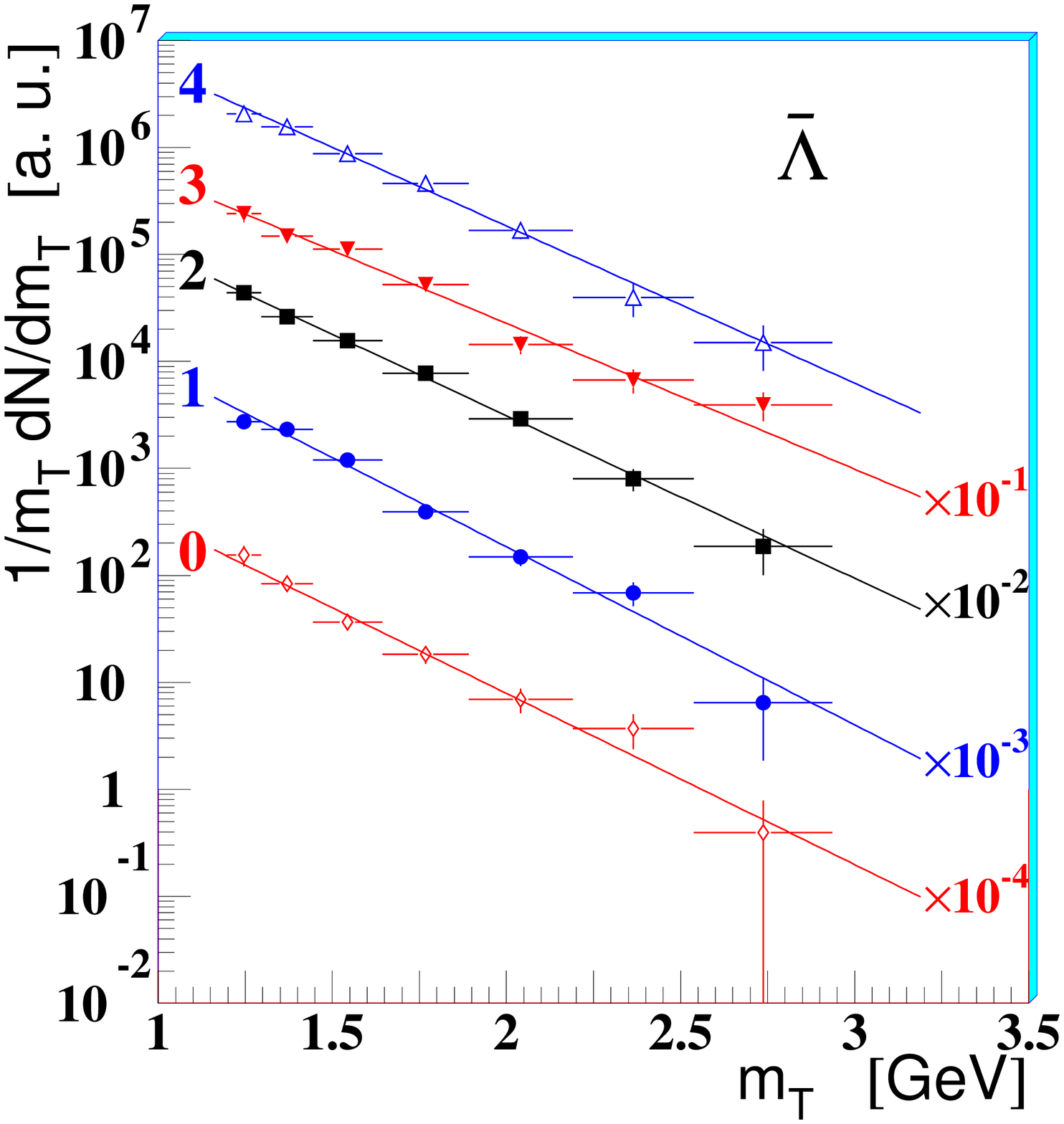}}\\
\resizebox{0.94\textwidth}{!}{%
\includegraphics{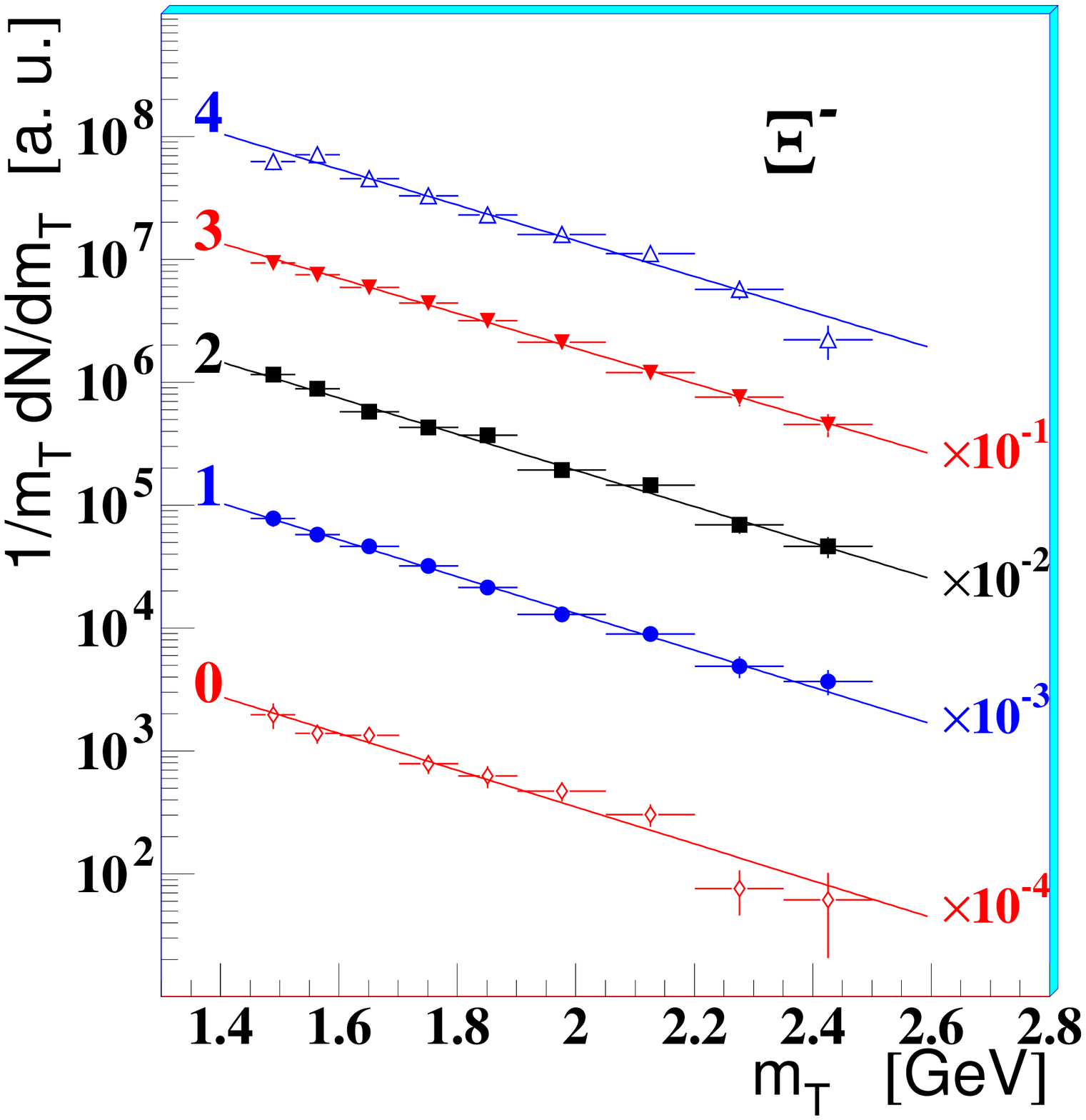}
\includegraphics{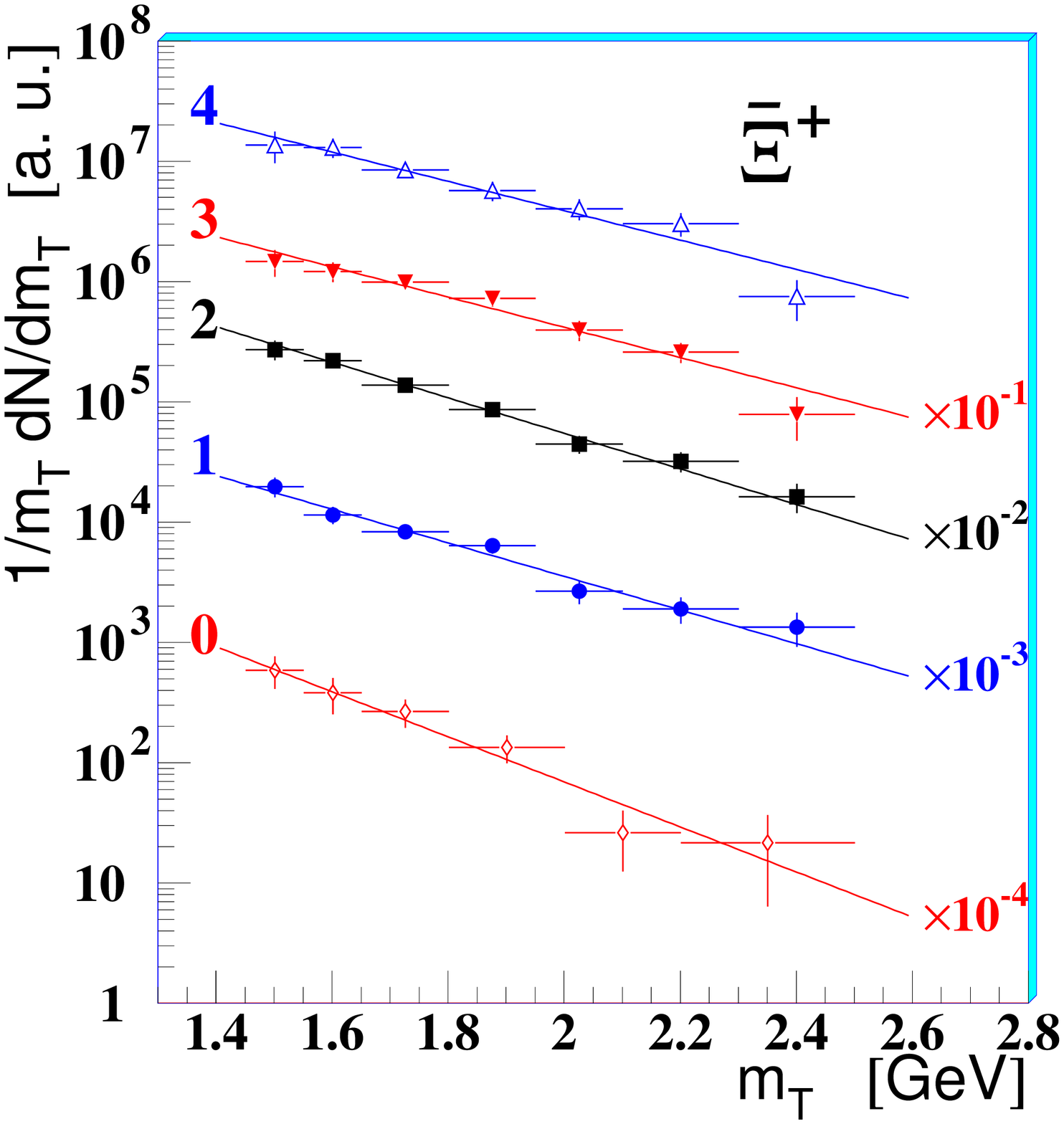}
\includegraphics{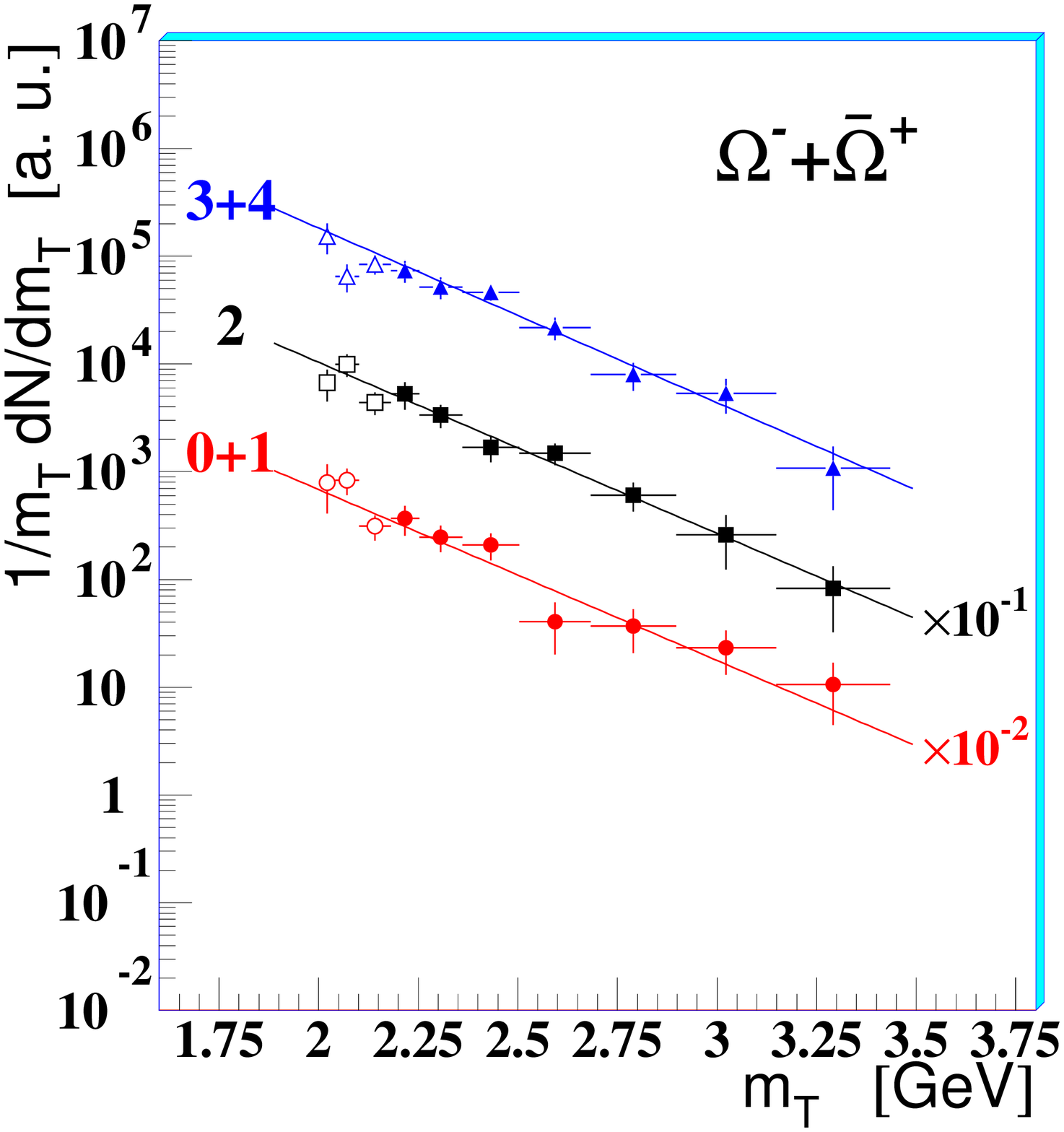}}
\caption{\rm  Transverse mass spectra of the strange particles for
              the five centrality classes of figure~\ref{fig:multiplicity}.
              For each species, class $4$\ is displayed uppermost and
              class $0$\ lowermost. In the \PgOm+\PagOp\
              spectra, the open points have not been used in the best fit calculation.}
\label{fig:msd_spectra}
\end{figure}
In order to reduce the statistical errors, the $\Omega^-$\ and  
$\overline\Omega^+$\ samples have been divided into only three  
centrality classes instead of five: $0+1$, $2$\ and $3+4$.    
The $1/m_{\tt T} \, dN/dm_{\tt T} $\ spectra are shown 
in figure~\ref{fig:msd_spectra} together with the   
exponential functions having inverse slopes as obtained from 
the maximum likelihood fit. The numerical values of the inverse  
slopes are given in table~\ref{tab:InvMSD}.   
\begin{table}[h]
\caption{Inverse $m_{\tt T}$\ slopes (MeV) of the strange particle distributions 
         for the five Pb-Pb centrality classes, and 
	 for p-Be and p-Pb interactions~\cite{Slope-p}.  
	 Only statistical errors are shown. In Pb-Pb, systematic errors are 
	 assumed to be independent of centrality.   
\label{tab:InvMSD}}
\begin{center}
\begin{tabular}{|c|c|c|cc|c|cc|} \cline{4-8}
      \multicolumn{3}{c}{ }    & \multicolumn{5}{|c|}{Pb-Pb} \\
\hline
      &  p-Be   &   p-Pb    &    0      &    1      &    2      &    3    &    4   \\ \hline
\PKzS &$197\pm4$& $217\pm6$ & $239\pm15$ & $239\pm8$ & $233\pm7$ & $244\pm8$ & $234\pm9$ \\
\PgL  &$180\pm2$& $196\pm6$ & $237\pm19$ & $274\pm13$ & $282\pm12$ & $315\pm14$ & $305\pm15$ \\
\PagL &$157\pm2$& $183\pm11$ & $277\pm19$ & $264\pm11$ & $283\pm10$ & $313\pm14$ & $295\pm14$ \\
\PgXm &$202\pm13$&$235\pm14$ & $290\pm20$ & $290\pm11$ & $295\pm9 $ & $304\pm11$ & $299\pm12$ \\
\PagXp&$182\pm17$&$224\pm21$ & $232\pm29$ & $311\pm23$ & $294\pm18$ & $346\pm28$ & $356\pm31$ \\
\PgOm+
\PagOp&$169\pm40$& $334\pm99$ & \multicolumn{2}{c|}{$274\pm34$} & $274\pm28$ &
        \multicolumn{2}{|c|}{$268\pm23$} \\
\hline
\end{tabular}
\end{center}
\end{table}
An increase of the inverse slopes with the centrality is observed in Pb-Pb  
for \PgL, \PagXp\ 
and possibly also for \PagL. 
\newline
Inverse slopes for p-Be and p-Pb collisions as measured by 
WA97~\cite{Slope-p} are also given in table~\ref{tab:InvMSD}. In central  
and semi-central Pb-Pb collisions (i.e. classes 1 to 4) one observes 
baryon-antibaryon symmetry in the shapes of the spectra:  
this can be seen in the values of the inverse slopes which are compatible for 
each hyperon and its antiparticle. 
This symmetry is not observed 
in p-Be 
collisions.  
\newline
The similarity of baryon and antibaryon $m_{\tt T}$\ slopes observed in Pb-Pb  
suggests that strange baryons and antibaryons are produced by a  similar   
mechanism in nuclear collisions.   
\section{Blast-wave description of the spectra}
%
In this section we use the statistical hadronization model of 
reference~\cite{BlastRef} to describe the strange particle spectra  
discussed above.   
The model assumes that particles decouple from a system in local thermal  
equilibrium with temperature $T$, which expands both longitudinally 
and in the transverse direction;    
the longitudinal expansion is taken to be boost-invariant  
and the 
transverse expansion is defined in terms of a transverse velocity 
field. Finally, the statistical distributions are   
approximated by the Boltzmann distribution.  
\newline
The blast-wave model predicts a double differential cross-section    
for a particle species $j$\   
of the form:  
\begin{equation}
\frac{d^2N_j}{m_{\tt T} dm_{\tt T} dy} 
    = \mathcal{A}_j  \int_0^{R_G}{ 
     m_{\tt T} K_1\left( \frac{m_{\tt T} \cosh \rho}{T} \right)
         I_0\left( \frac{p_{\tt T} \sinh \rho}{T} \right) r \, dr}
\label{eq:Blast}
\end{equation}
where $\rho(r)=\tanh^{-1} \beta_{\perp}(r)$, $K_1$\ and $I_0$\ are two 
modified Bessel functions and $\mathcal{A}_j$\ is a normalization 
constant.  
With respect to a cylindrical reference system ($r$,$\phi$,$z$,$t$), 
the freeze-out hypersurface is constrained by $0 \le r \le R_G$,   
$ 0 \le \phi \le 2\pi$ and $\partial t_f / \partial r = 0$;   
the last condition assumes that the particles decouple suddenly   
from  the whole transverse profile of the 
fireball at time $t_f$. 
In these expressions $R_G$\ is the transverse geometric 
radius of the source.  
\newline
In case of a peripheral collision, the azimuthal symmetry is evidently broken; 
however it is recovered when considering the $m_{\tt T}$\ spectrum of particles 
accumulated over 
many 
events with random impact parameters, which is 
our approach.   
\newline
The transverse velocity field $\beta_{\perp}(r)$\ can be parametrized 
according to a power law: 
\begin{equation}
\beta_{\perp}(r) = \beta_S \left[ \frac{r}{R_G} \right]^{n}  
  \quad \quad \quad r \le R_G
\label{eq:profile}
\end{equation}  
With  
this type of profile the numerical value of $R_G$\ does not 
influence the shape of the spectra but just the absolute  normalization 
(i.e. the $\mathcal{A}_j$\ constant). 
Once the transverse flow profile (i.e. equation~\ref{eq:profile}) has been fixed,  
the shape of each spectrum is determined by the temperature,
the velocity of the transverse expansion and the mass of the particle.   
The parameters which can be extracted from a fit of equation~\ref{eq:Blast} to 
the experimental spectra are thus the thermal freeze-out 
temperature $T$\ and the surface transverse flow velocity $\beta_S$. 
In order to compare 
results from different profile hypotheses, 
corresponding to different values of the exponent $n$\ in equation~\ref{eq:profile}, 
the average transverse flow velocity can be used instead. 
Assuming a uniform particle density, the average\footnote[2]{A more 
sophisticated averaging can be 
achieved by incorporating not only the transverse geometry of the model 
but also the phase space density of particles~\cite{Nonso}.  
According to this definition, $<\beta_{\perp}>$\ is also  
a function of $T$\ and it differs from the  
values calculated according to equation~\ref{eq:averageB} 
by 2\% if $n=1/2$, by 5\% if $n=1$\ and by 10\% if $n=2$; obviously 
for $n=0$, $<\beta_{\perp}>=\beta_S$\ independently of the average definition.}  
is~\cite{Nonso}: 
\begin{equation}
<\beta_{\perp}> = \frac{2}{2+n}  \beta_S
\label{eq:averageB}
\end{equation}
%
\subsection{Global fits to the spectra using different flow velocity profiles}
Preliminary NA57 results of the blast-wave model were 
presented~\cite{Bruno2} using a constant ($n=0$) velocity profile, 
for the purpose of comparison  
with the NA49 analysis~\cite{BlastNA49}.   
In the following, we will adopt primarily the linear ($n=1$)  
$r$-dependence of the transverse flow velocity. In fact,  
a constant  
flow profile description of the transverse expansion  
is internally inconsistent, because the outermost fluid elements of  
the expanding fireball should have travelled faster than  
the innermost ones: $\beta(r_1) > \beta(r_2)$\ if $r_1 > r_2$.  
On the other hand, the quadratic profile ($n=2$) is disfavoured by data, 
as will be shown 
below.  
We also consider the $n=1/2$\ profile which has  
been suggested  
to resemble closely  
the solution of the full hydro-dynamical calculation~\cite{Urs}.  

\begin{figure}[hbt]
\centering
\resizebox{0.96\textwidth}{!}{%
\includegraphics{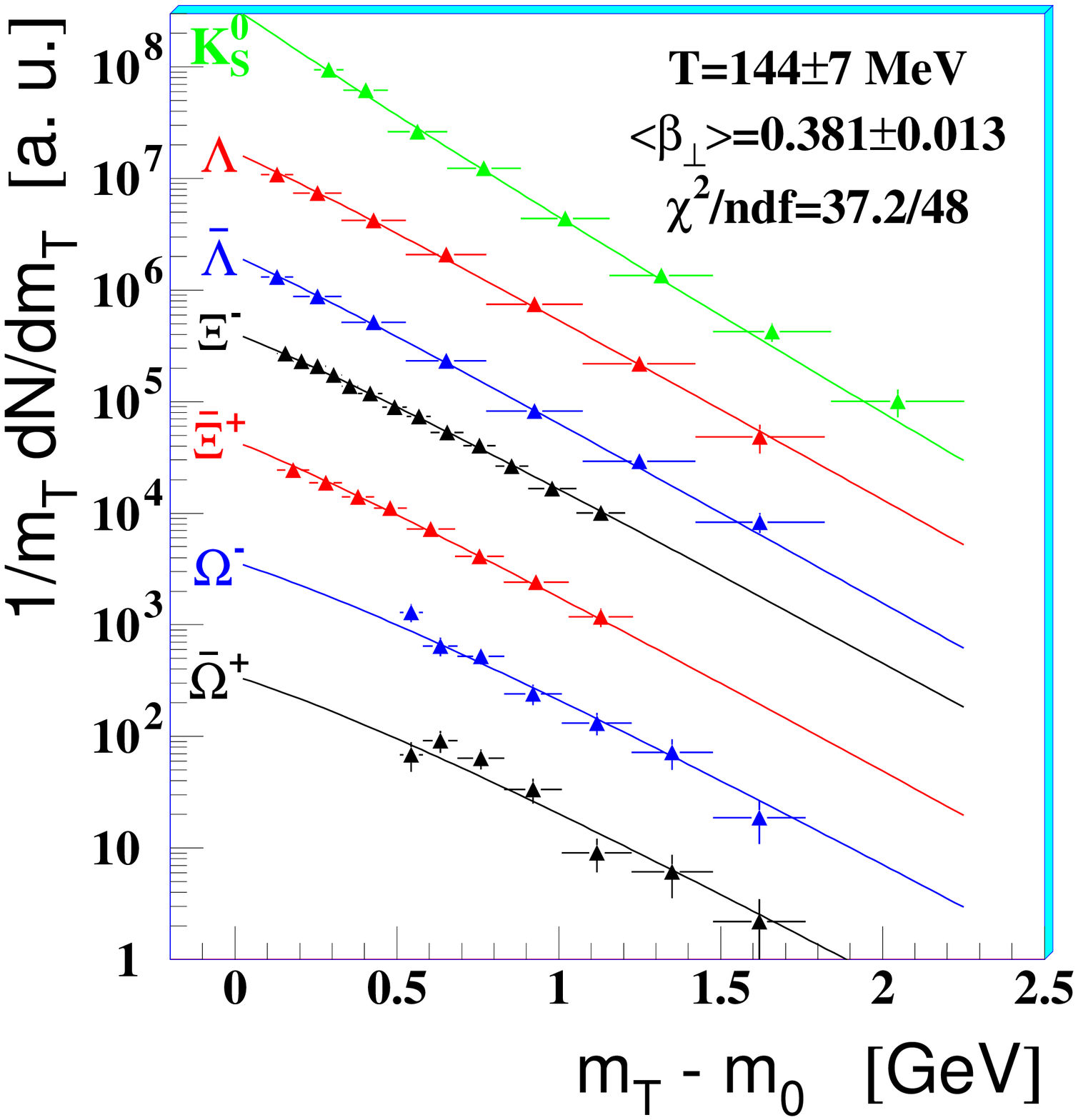}
\includegraphics{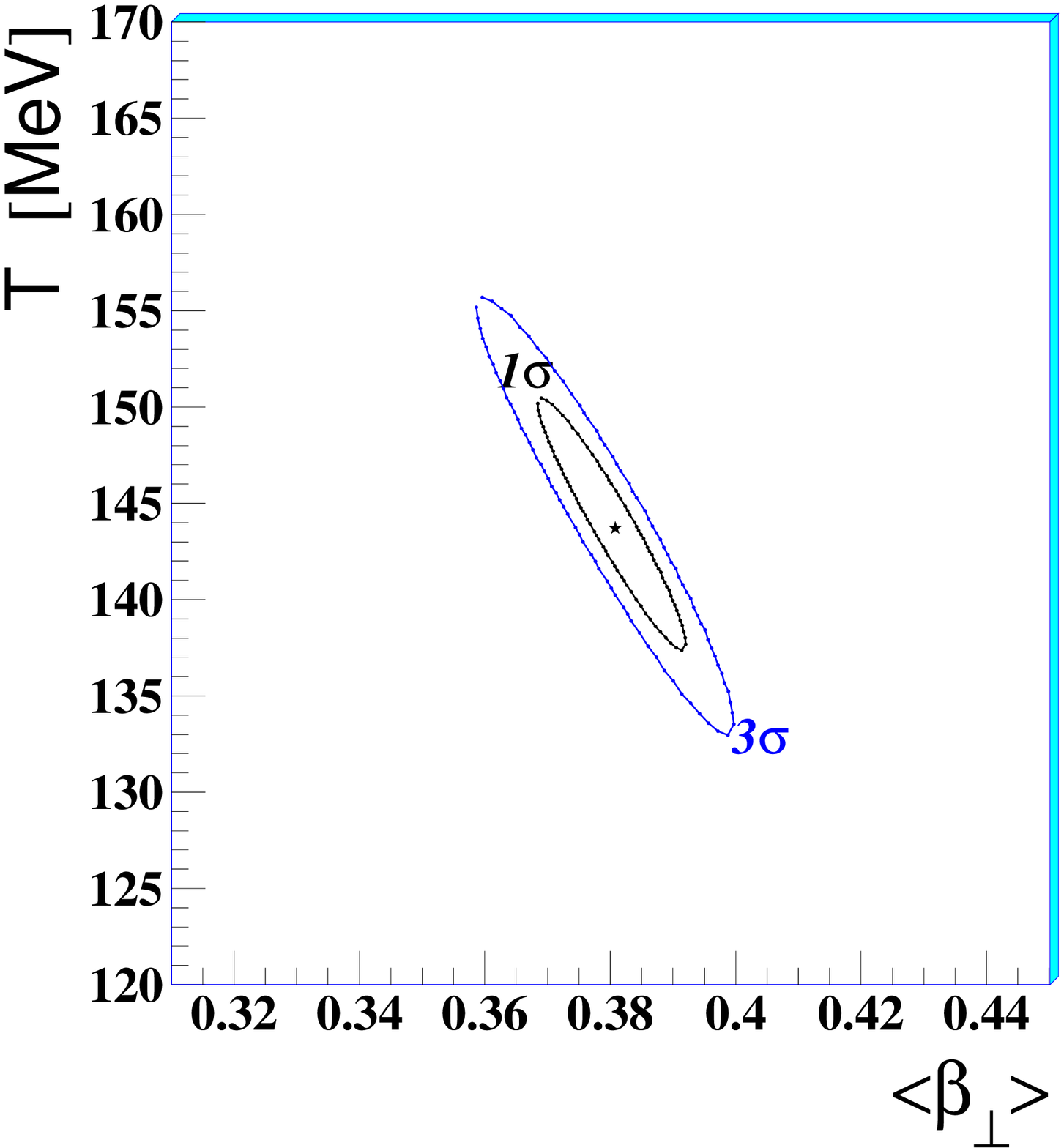}}
\caption{\rm Blast-wave fit to the spectra with a linear profile (left);  
             corresponding 1$\sigma$\ and 3$\sigma$\ confidence level contours 
	     in the $<\beta_{\perp}>$--$T$\ plane (right).}  
\label{fig:Blast_lin}
\end{figure}

The global fit of equation~\ref{eq:Blast} with $n=1$\ 
to the data points of all the measured strange particle spectra 
is shown in figure~\ref{fig:Blast_lin}; 
it successfully describes all the distributions 
with $\chi^2/ndf=37.2/48$, yielding the following values for the two parameters  
$T$\ and $ <\beta_\perp>$:  
\[ \fl
 T = 144 \pm 7 {\tt (stat)} \pm 14 {\tt (syst)} {\rm MeV} \, , \quad 
 <\beta_\perp>=0.381 \pm 0.013 {\tt (stat)} \pm 0.012 {\tt (syst)} \; .
\nonumber 
\]
The $T$\ and $<\beta_\perp>$\ parameters are found to be statistically 
anti-correlated, as can be seen from the confidence level contours shown    
in figure~\ref{fig:Blast_lin}.  Conversely, the systematic errors on $T$\ 
and $<\beta_{\perp}>$\  are found to be correlated  
and they are estimated to be $10\%$\ and $3\%$, respectively.  

\begin{figure}[hbt]
\centering
\resizebox{0.54\textwidth}{!}{%
\includegraphics{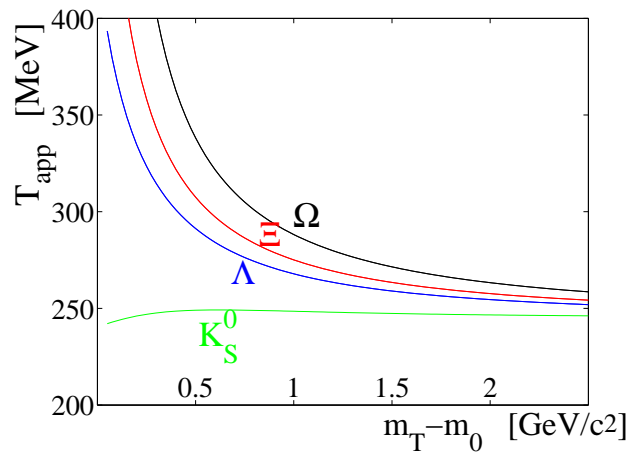}}
\caption{\rm Local inverse slopes according to the  
             blast-wave model using the $n=1$\ transverse profile.}  
\label{fig:Blast_der}
\end{figure}

The results of the fits with all the  four profile hypotheses are  
given in table~\ref{tab:profile}.  
\begin{table}[htb]
\caption{Results of the blast-wave model fit using different  
         transverse flow profiles. The quoted errors are statistical.  
	 The systematic errors on the temperature and on the velocities 
	 are estimated to be about $10\%$\ and $3\%$, respectively,  
	 for all the four profiles.  
\label{tab:profile}}
\begin{center}
\begin{tabular}{|c||c|c|c|c||}
\hline
                    & {\bf $n=0$}     & {\bf $n=1/2$}   & {\bf $n=1$}     & {\bf $n=2$}   \\ \hline 
 $T$\ (MeV)         &  $158 \pm 6$    & $ 152\pm6 $     & $ 144\pm7 $     & $151 \pm 11$    \\ \hline
 $\beta_{S}      $  & $0.396\pm0.015$ & $0.493\pm0.016$ & $0.571\pm0.019$ & $0.633\pm0.028$ \\ \hline
 $<\beta_{\perp}>$  & $0.396\pm0.015$ & $0.394\pm0.013$ & $0.381\pm0.013$ & $0.316\pm0.014$ \\ \hline
 $\chi^2/ndf     $  & $ 39.6/48 $     &  $ 36.9/48 $    & $ 37.2/48 $     & $68.0/48 $      \\ \hline  
\hline
\end{tabular}
\end{center}
\end{table}  
\noindent 
The three profiles $n=0$, $n=1/2$\ and $n=1$\ yield    
similar values of the freeze-out temperatures and of the  average transverse 
flow velocities, with good values of $\chi^2/ndf$. 
The quadratic profile is disfavoured by our data.  
For $m_{\tt T}$\ sufficiently 
larger than the 
particle mass $m_0$, the inverse slope $T_{app}$\ 
is expected to be 
independent of $m_0$~\cite{BlastRef2}.  
In the non-relativistic regime ($p_{\tt T} < m_0$) $T_{app}$\ is    
expected to increase with the rest mass of the particle;   
e.g. in  reference~\cite{Heinz} it has been derived that   
$T_{app} \approx T + \frac{1}{2} m_0 <\beta_\perp>^2$:
the larger the mass of the particle the more pronounced the flattening 
of the spectra at low $m_{\tt T}$.  
The deviations from exponential behaviour are larger for large 
flows and 
at low $m_{\tt T}$.  
Due to its light mass, the \PKzS\ spectrum is predicted by the 
model, and indeed observed, to be steeper at low $m_{\tt T}$.   
Such behaviour is illustrated in figure~\ref{fig:Blast_der} where we 
compute the ``local''  inverse slope,  i.e. the inverse slope  
$T_{app}$\ of the exponential function $\exp(-m_{\tt T}/T_{app})$\  
which is tangent to the blast-wave curve, as a function of $m_{\tt T} - m_0$.   
Pions, which in principle would be an even better probe for
this behaviour, are unfortunately affected by resonance decays (at low $p_{\tt T}$)
which cannot be neglected~\cite{Sollfrank,HeavyIonPhys}.
%
%

The particles have been divided into two groups --- those which share  
valence quarks with the nucleons and those which do not  --- since 
it is known that the particles of the two groups may exhibit different 
production features. 
The fit procedure has been repeated separately for the two groups 
($n=1$\ case) and the results are given in table~\ref{tab:Blast1}.  
\begin{table}[h]
\caption{Thermal freeze-out temperature ($T$), surface ($\beta_S$) 
and average ($<\beta_\perp>$) 
transverse flow velocity in the full centrality range,   
assuming a linear transverse profile ($n=1$). 
The first error is statistical, the second one systematic.  
\label{tab:Blast1}}
\begin{center}
\begin{tabular}{|c||c|c|c|c||}
\hline
particles & ${\bf T}$ (MeV) & $\beta_S$ & $<{\bf \beta_\perp}>$ & $\chi^2/ndf$ \\ \hline
{\bf $K_S^0$}, {\bf $\La$},  {\bf $\Xi^-$} &
$146 \pm 8 \pm 14 $ & $0.564\pm0.023\pm0.018$ & $ 0.376 \pm 0.015 \pm 0.012 $ & $ 18.1/23 $ \\ \hline
{\bf $\Al$}, {\bf $\overline\Xi^+$}, {\bf $\Omega^-$}, {\bf $\overline\Omega^+$} &
$ 130\pm28 \pm 14 $ & $0.604\pm0.048\pm0.018$ & $ 0.403 \pm 0.032 \pm 0.012 $ & $ 18.5/23 $ \\
\hline
\end{tabular}
\end{center}
\end{table}
\noindent
They suggest common freeze-out conditions for the two groups. 
Since the interaction cross-sections for
the particles of the two groups are quite different, this finding would suggest
limited importance of final state interactions  
(i.e. a rapid thermal freeze-out) and  
a similar production mechanism.  
A similar conclusion concerning the evolution of the system was reached  
from the study of the HBT correlation functions of 
negative pions~\cite{HBTNA49,HBTpaper,HBTCERES}.  

\begin{figure}[hbt]
\centering
\resizebox{0.54\textwidth}{!}{%
\includegraphics{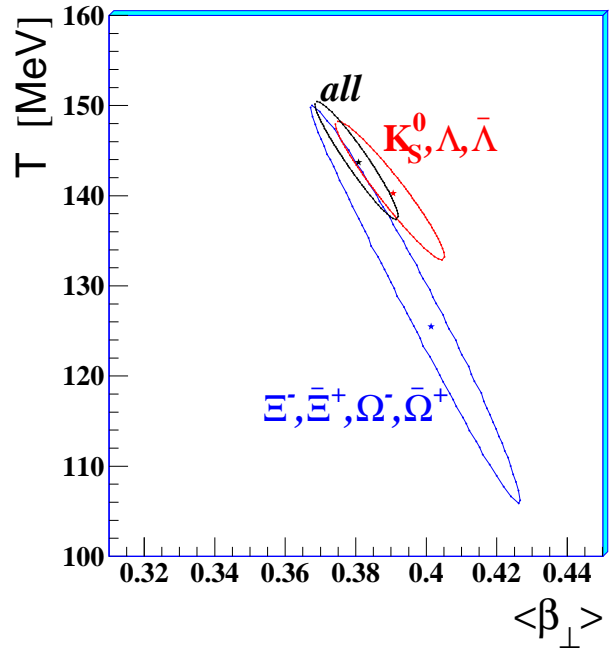}
}
\caption{
         The 
	 thermal freeze-out temperature vs average transverse flow for the 
         blast-wave fits ($n=1$\ profile) 
	 to multiply and singly strange particle spectra.   
	 The 1$\sigma$\ contours are shown, with the marker indicating 
	 the optimal fit location. 
	 The global fit contour ({\em all}) is also shown.   
         }
\label{fig:Strange}
\end{figure}

It has been suggested~\cite{Hecke}, based on WA97 results on the hyperon 
$m_{\tt T}$\  
slopes compatible with those discussed in section 4, that the thermal 
freeze-out occurs earlier for $\Omega$\ and possibly for $\Xi$\ than for 
particles of strangeness 
$0$, due to the low scattering  cross-sections for $\Omega$ and $\Xi$.  
The 1$\sigma$\ contours of the separate blast wave fits for singly  
and multiply strange particles are shown in figure~\ref{fig:Strange}. 
Both 
groups of particles are compatible 
  with the global-fit determination.   
However, the fit for the multiply strange particles is statistically  
dominated by the $\Xi$;  in fact the $\Xi+\Omega$\  
contour remains essentially unchanged when fitting the $\Xi$\ alone.  
Therefore we can only conclude   
that the $\Xi$\ undergoes a thermal freeze-out which is compatible with 
that of $\Lambda$\ and $\PKzS$.  
\newline
Due to the lower statistics of the $\Omega$,  
from its spectrum alone it was not possible to extract significant values  
for the two freeze-out parameters, as can be done for the $\Xi$.  
Any possible deviation of the $\Omega$\ 
from the observed common freeze-out undergone 
by $\PKzS$, $\Lambda$\ and $\Xi$\   
can only be inferred from the integrated information of the 
$\Omega$\ spectrum, i.e. from its inverse slope.  
\noindent
\begin{figure}[hbt]
\centering
\resizebox{0.54\textwidth}{!}{%
\includegraphics{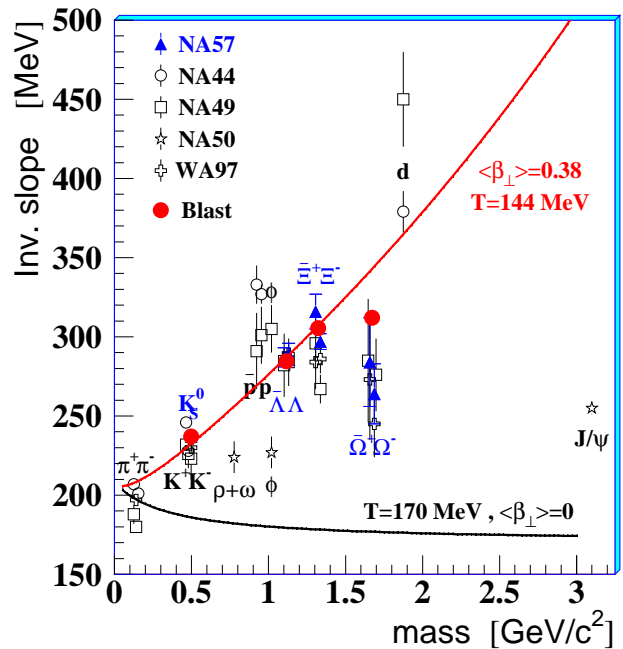}}
\caption{Prediction of the blast-wave model for
         inverse slopes (see text for details).}
\label{fig:BlastPred}
\end{figure}

%
In figure~\ref{fig:BlastPred} we plot the inverse slopes of 
figure~\ref{fig:Inv_mass}  
superimposing the blast-wave model results.   
The full lines represent the blast-wave 
calculations when fitting the $m_{\tt T}$\ spectrum of a particle, as a function 
of mass,  
in the common range $ 0.05 < m_{\tt T} - m_{0}  < 1.50$\ GeV/$c^2$, for two different  
freeze-out conditions: 
absence of transverse flow ($<\beta_{\perp}>=0$) and our best fit determination.  
The inverse slope is also a function of the $m_{\tt T} - m_{0}$\ 
range where the fit is performed. This dependence is stronger for 
heavier particles (see figure~\ref{fig:Blast_der}). Therefore we have also 
computed the blast-wave inverse slopes of $\PKzS$, $\PgL$, $\Xi$\ and $\Omega$\  
spectra in the $m_{\tt T} - m_{0}$\ ranges of NA57  
(closed circles in figure~\ref{fig:BlastPred}). 
The measured inverse slopes of $\Omega$\ show a deviation  
from the global trend of the other strange particles.  
\subsection{Centrality dependence} 
\begin{figure}[hbt]
\centering
\resizebox{0.50\textwidth}{!}{%
\includegraphics{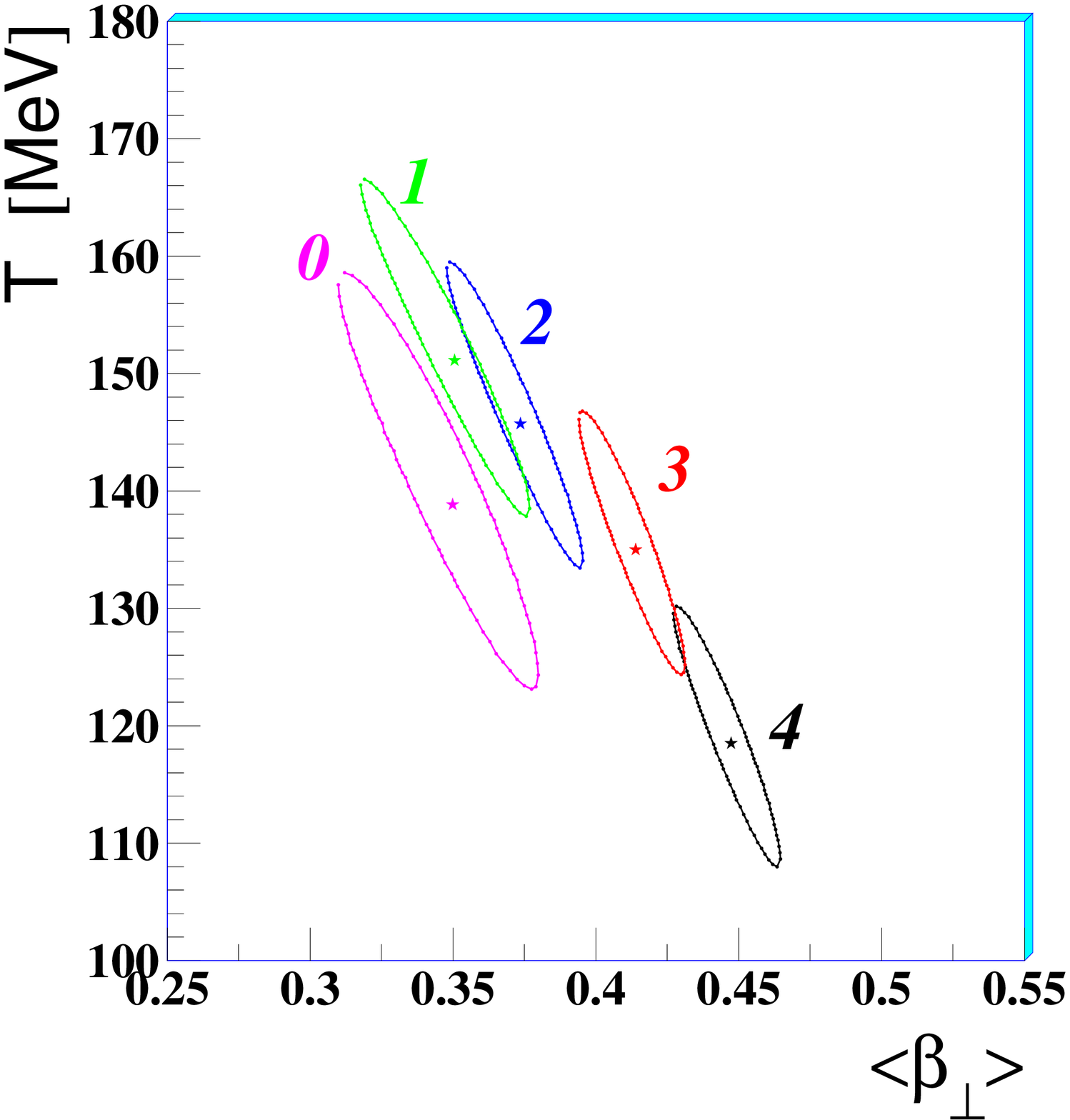}}
\caption{\rm The 1$\sigma$\ confidence level contours from
               fits in each centrality class.} 
\label{fig:cont_msd}
\end{figure}

We have  performed the global fit to the spectra  
for each of the five centrality classes defined in section~3.  
In figure~\ref{fig:cont_msd} we show the $1\sigma$\ confidence level contours 
in the $<\beta_\perp>$--$T$\ plane as 
obtained for the $n=1$\  profile.   
The numerical values of the fit parameters are  
given in table~\ref{tab:BlastMSD}.  
\begin{table}[h]
\caption{Thermal freeze-out temperature and   
transverse flow velocity in the five centrality classes, 
assuming a linear transverse profile ($n=1$). Only statistical errors are 
quoted; the systematic errors on $T$\ and $<\beta_\perp>$\  
are estimated to be $10\%$\ and $3\%$, respectively.  
We also give the blast-wave determinations of $T$\ and $<\beta_{\perp}>$\  by 
the NA49 experiment in Pb-Pb collisions at 158 $A$\ 
GeV/$c$~\cite{BlastNA49}, and by  
the PHENIX and STAR experiments in Au-Au collisions at  
$\sqrt{s_{NN}}=130$\ GeV~\cite{PHENIX130,STAR130}.  
Finally,  we quote the measurements by the NA49~\cite{HBTNA49}, WA97~\cite{HBTpaper} 
and CERES~\cite{HBTCERES} experiments which use HBT correlation  
combined with information from $h^-$\ and/or deuterium single particle spectra.  
\label{tab:BlastMSD}}
\begin{center}
\footnotesize{
\begin{tabular}{|l|c|c|l|c|c|c|}
\hline
Exp. & $\sigma/\sigma_{inel}$\ \% & $\sqrt{s_{NN}}$\ &\multicolumn{1}{c|}{technique} & 
                              ${\bf T}$ (MeV)&$<{\bf \beta_\perp}>$&$\chi^2/ndf$\\ \hline
NA57& 40-53 &$17.3$& this analysis, $n=1$& $139\pm25$ & $0.35\pm0.04$ &  $33.2/35$ \\ 
NA57& 23-40 &$17.3$& this analysis, $n=1$& $151\pm15$ & $0.35\pm0.03$ &  $42.3/42$ \\ 
NA57& 11-23 &$17.3$& this analysis, $n=1$& $146\pm17$ & $0.37\pm0.03$ &  $41.3/43$ \\ 
NA57& 4.5-11&$17.3$& this analysis, $n=1$& $135\pm14$ & $0.41\pm0.02$  & $31.8/41$ \\ 
NA57& 0-4.5 &$17.3$& this analysis, $n=1$& $118\pm13$ & $0.45\pm0.02$  & $52.7/43$ \\ 
NA49& 0-5,10,20 &$17.3$&\PKp,\Pp,\PgL,\PgXm,\PgOm,  $n=0$ & $127\pm4$ & $0.48\pm0.01$  &$118/43$ \\
NA49& 0-5,10,20 &$17.3$&\PKm,\Pap,$\phi$,\PagL,\PagXp,\PagOp, $n=0$ & $122\pm2$ & $0.48\pm0.01$ & $46/41$  \\
PHENIX& 60-92 &$130$&\Pgppm,\PKpm,\Pp,\Pap, $n=1$ & $161^{+19}_{-12}$ & $0.16^{+0.16}_{-0.2}$  & $36.3/40$\\
PHENIX& 30-60 &$130$&\Pgppm,\PKpm,\Pp,\Pap, $n=1$ & $140\pm4$ & $0.39\pm0.01$  & $68.9/40$\\
PHENIX& 15-30 &$130$&\Pgppm,\PKpm,\Pp,\Pap, $n=1$ & $134\pm2$ & $0.43\pm0.01$  & $36.2/40$\\
PHENIX& 5-15 &$130$&\Pgppm,\PKpm,\Pp,\Pap, $n=1$ & $125\pm2$ & $0.46\pm0.01$  & $34.7/40$\\
PHENIX& 0-5 &$130$&\Pgppm,\PKpm,\Pp,\Pap, $n=1$ & $121\pm4$ & $0.47\pm0.01$  & $34.0/40$\\
STAR& 0-10 &$130$&\PgXm, \PagXp,  $n=1/2$   & $ 182\pm29$ & $0.42\pm0.06$ & $13/15$ \\
STAR& 0-10 &$130$&\Pgp,\PK,\Pp,\PgL,  $n=1/2$ & $103\pm7$ & $0.57\pm0.01$ &   \\
\hline
NA49& 0-3 &$17.3$& HBT + $h^-$ + $d$ & $   120\pm12 $    & $0.55\pm0.12$\\
WA97& 25-40 &$17.3$& HBT + $h^-$ & $140^{+26}_{-13}$ & $0.30^{+0.09}_{-0.16}$ \\
WA97& 12-25 &$17.3$& HBT + $h^-$ & $121^{+15}_{-11}$ & $0.47^{+0.07}_{-0.10}$ \\
WA97& 5-12 &$17.3$& HBT + $h^-$  & $117^{+16}_{-11}$ & $0.48^{+0.08}_{-0.11}$ \\
WA97& 0-5 &$17.3$& HBT + $h^-$   & $120^{+15}_{-11}$ & $0.46^{+0.07}_{-0.10}$\\ 
CERES& 15-30 &$17.3$& HBT &\footnotesize{fixed at} $120$ & $0.49^{+0.06}_{-0.06}$   \\
CERES& 10-15 &$17.3$& HBT &\footnotesize{fixed at} $120$ & $0.53^{+0.06}_{-0.05}$   \\
CERES& 5-10 &$17.3$& HBT &\footnotesize{fixed at} $120$ & $0.46^{+0.04}_{-0.04}$   \\
CERES& 0-5 &$17.3$& HBT &\footnotesize{fixed at} $120$ & $0.55^{+0.03}_{-0.03}$   \\ \cline{1-6}
\end{tabular}
}
\end{center}
\end{table}
The observed trend is as follows:  
the more central the collisions the larger the transverse collective 
flow and the lower the final thermal freeze-out temperature.  
The higher freeze-out temperature for peripheral collisions may be interpreted 
as the result of an earlier decoupling of the expanding system.  
\newline
A similar centrality dependence of the freeze-out parameters has been observed  
by the PHENIX experiment in Au-Au collisions   
at $\sqrt{s_{NN}}=130$\ GeV from the blast-wave analysis 
of  \Pgppm, \PKpm, \Pp\ and \Pap\ spectra~\cite{PHENIX130}.   
In particular,   
PHENIX has measured\footnote[2]{In the PHENIX analysis, statistical
and systematic errors are added in quadrature before the fit.} 
$T=121 \pm 4$\ MeV, $<\beta_\perp> = 0.47 \pm 0.01$\ 
with $\chi^2/ndf=34.0/40$\ for the most central 5\% of the Au-Au cross-section and   
$T=140 \pm 4$\ MeV, $<\beta_\perp> = 0.39 \pm 0.01$\  with $\chi^2/ndf=68.9/40$\ 
for peripheral collisions (30--60\% centrality,   
which corresponds roughly to the NA57 class 0).   
For the same colliding system, 
the STAR experiment  
has also determined~\cite{STAR130} the thermal freeze-out for central 
collisions (see table~\ref{tab:BlastMSD}):   
STAR results would suggest a rather earlier decoupling of   
the multi-strange $\Xi$\ with respect to non strange and singly strange   
particles at RHIC energy.   
\newline
We also quote the result of our fit 
for the most central class, when the $n=0$\  
profile is assumed: $T=134 \pm 10 \pm 13 $\ MeV,   
$<\beta_\perp>=0.457\pm0.025 \pm 0.014$\ and 
$\chi^2/{\rm ndf} =47.2/43$. 
This result is found to be consistent within two standard deviations 
with that from a similar analysis performed by NA49\footnote{
The NA49 centrality ranges are 0-5\% for $K^{\pm}$\ and $\phi$, 0-10\% 
for \Pgp, \PgL\ and $\Xi$, 0-20\% for $\Omega$.} 
assuming the constant profile hypothesis~\cite{BlastNA49}.  
\newline
In table~\ref{tab:BlastMSD} we also make a comparison with results from  
different measurements which exploit the $p_{\tt T}$\  
dependence of the HBT correlation functions of negative pions~\cite{HBTNA49,HBTpaper,HBTCERES}.  
The thermal freeze-out temperatures and transverse flow velocities are found to be in good 
overall agreement with those extracted from the blast-wave fits.  
\section{Conclusions}
We have analysed the transverse mass spectra of 
high statistics, high purity samples of \PKzS,  \PgL, $\Xi$\ and $\Omega$\   
particles produced in Pb-Pb collisions at 158 $A$\ GeV/$c$\ over a wide range 
of collision centrality (the most central 53\% of the Pb-Pb inelastic   
cross-section).  

The inverse slopes agree with those measured by WA97. 
For each hyperon species, the $m_{\tt T}$\ slope is found to be in good agreement 
with that of the antiparticle.    
An increase of the inverse slopes with the collision centrality   
is observed for \PgL\ and \PagXp.     

The analysis of the transverse mass spectra at 158 $A$\ GeV/$c$\ 
in the framework of the blast-wave  model 
suggests  that after a central 
collision the system expands explosively; 
the system then freezes-out when the temperature is of the order 
of 120 MeV with an average transverse  flow   
velocity of about one half of the speed of light.  
Particles with and without valence quarks in common with the nucleon 
appear to have a similar behaviour.   
The inverse slope of the $\Omega$\ particle deviates from the prediction of  
the blast-wave model tuned on the other strange particles (\PKzS, \PgL\ and $\Xi$).    
Thermal freeze-out conditions (i.e. temperature and transverse flow velocity)  
are found to be similar to those measured at RHIC energy for singly strange and   
non strange particles, in both central and peripheral collisions; conversely the  
temperature measured for the multiply strange $\Xi$\  
is significantly larger at RHIC.   
Finally, the  results on the centrality dependence of the  
expansion dynamics  
indicate that with increasing centrality  
the transverse flow velocity increases and the final temperature decreases.  
%
%
\ack
We are grateful to U~Heinz, J~Rafelski, B~Tom\'{a}\v{s}ik and U~A~Wiedemann
for useful comments and fruitful discussions.   

\end{document}